\documentclass[11pt,letter]{article}
\pdfoutput=1 

\usepackage{jheppub}
\usepackage{amssymb,amsmath,amsthm,graphicx}
\usepackage{graphics, color}
\usepackage{latexsym}
\usepackage{bm}
\usepackage[hang,flushmargin]{footmisc}
\usepackage{epsfig}
\usepackage{textcomp}
\usepackage{booktabs}
\usepackage{array}
\usepackage{caption}
\usepackage{tikz}
\usetikzlibrary{positioning}
\usetikzlibrary{calc}
\hypersetup{
 pdfauthor={William L. Boyce, Jorge  E. Santos},
 citecolor=black,linkcolor=black,urlcolor=black}
\usepackage[utf8]{inputenc}
\allowdisplaybreaks[1]

\def \dd {\mathrm{d}}
\def \ii {\mathrm{i}}
\def\ie{{\it i.e.} }
\def\eg{{\it e.g.} }

\def \Re {\mathrm{Re}}
\def \Im {\mathrm{Im}}

\begin{document}
\title{\centering{Amplified EFT corrections near exceptional QNMs}}
\author{William~L.~Boyce}
\author{and Jorge~E.~Santos}

\affiliation{DAMTP, Centre for Mathematical Sciences, University of Cambridge, Wilberforce Road, \\ Cambridge CB3 0WA, UK}
\emailAdd{wlb26@cam.ac.uk}
\emailAdd{jss55@cam.ac.uk}

\newcommand{\blue}{\color{blue}}
\newcommand{\red}{\color{red}}

\abstract{
We show that the ordinary expansion of EFT corrections to individual QNM frequencies breaks down at an exceptional QNM, and that these corrections become enhanced for near-exceptional QNMs. We give a brief argument for why this should generally happen, before numerically demonstrating this behaviour by calculating the $\alpha'^3$ corrections to QNMs of equal-spin Myers--Perry-AdS$_5$ black holes. This calculation corresponds to $\lambda^{-3/2}$ corrections to the poles of the retarded Green's function in the boundary CFT, where $\lambda$ is the 't Hooft coupling. Treating the coalescing pair together, we calculate the leading $\lambda^{-3/4}$ splitting at the uncorrected exceptional point and the $\lambda^{-3/2}$ displacement of the exceptional point in temperature and angular velocity. The background derivative expansion can remain valid while the nondegenerate expansion of individual frequencies fails. We discuss the implications for the full response and for other areas of black hole physics.
}

\maketitle

\section{Introduction}\label{sec: intro}
General Relativity (GR) is an immensely successful theory of physics, able to describe a wide range of gravitational phenomena on large scales. However, GR should only be treated as an effective theory of gravity and must be supplemented by higher-derivative corrections, which become important at shorter length-scales and larger curvatures. 

Despite this need to consider higher-derivative theories, we are limited by the fact that these theories generally are not known to admit a well-posed formulation when treated exactly (with exceptions when the equations of motion are second order \cite{Kovacs:2020pns,Kovacs:2020ywu,Davies:2021frz}), and furthermore there is an infinite-dimensional space of possible higher-derivative operators which could appear in the relevant Lagrangians. As such, it is instead sensible to work at length and curvature scales for which these higher-derivative corrections can be treated as perturbative. We then consider only the leading-order higher-derivative corrections to GR, which span a finite-dimensional vector space. In this way we can systematically describe all possible behaviours that a UV complete theory might have at leading order\footnote{EFTs can also describe higher-order corrections to GR. In this paper we restrict to first-order for simplicity.}, assuming no new light fields, while admitting a well-posed formulation \cite{Figueras:2024bba}. This is the effective field theory (EFT) approach which we follow in this paper. 

At first order in EFT, calculating the correction to a differentiable physical observable is equivalent to taking its derivative with respect to a coupling parameter at the GR limit. A relevant example of this is the higher-derivative string-theoretic corrections to type IIB SUGRA on AdS$_5 \times S^5$, which correspond to taking a derivative with respect to $\lambda^{-3/2}$ at $\lambda = \infty$, where here $\lambda$ is the 't Hooft coupling of the dual gauge theory. 

Treating EFT as a derivative poses a natural question: Can we find observables whose derivative with respect to this coupling blows up? If so, the coefficient of the first-order correction becomes arbitrarily large, and the ordinary expansion in the coupling is no longer uniform. In this paper, by a breakdown of the expansion we mean this failure for individual QNM frequencies. The EFT description of the spacetime geometry can remain valid, and a different expansion that treats a nearly degenerate pair together can still give a controlled leading prediction.

Enhanced corrections have already been observed for tidal forces of near-extremal black holes \cite{Horowitz:2023xyl,Horowitz:2024dch}, for quasinormal modes (QNMs) in the microcanonical ensemble as we approach extremality \cite{Boyce:2026rnn}, and for eikonal QNMs in the same limit \cite{Cano:2025mht,Cano:2025ejw}. It is perhaps not surprising that such behaviour occurs near extremal black holes, whose linearised dynamics are known to display more subtle behaviour, \eg the Aretakis instability \cite{Aretakis:2012ei}. It would be tempting to conclude from these studies that a singular expansion of black hole observables must occur in the near-extremal regime.

In this paper we will demonstrate that the ordinary expansion of QNM frequencies can fail even for \emph{sub}extremal black holes. This is traced back to a non-analyticity in the QNM spectrum of black holes---namely the phenomenon of exceptional QNMs, closely related to avoided crossings and QNM hysteresis. At an exceptional QNM, two frequencies and their corresponding eigenfunctions coalesce. In a generic two-parameter family, such degeneracies, when present, occur at isolated points after tuning two real dimensionless parameters. Exceptional QNMs have been observed for a wide range of families of black holes, such as Kerr--Newman \cite{Dias:2021yju,Dias:2022oqm} and Kerr--de Sitter \cite{Rossi:2026als}. The same effect can also arise for \eg a one-parameter family of black holes and one-parameter family of theories, such as a massive scalar field around the Kerr black hole \cite{Cavalcante:2024swt,Cavalcante:2024kmy,Cavalcante:2025abr}. Exceptional QNMs can also be seen when investigating black brane QNMs with \emph{complex} spatial momenta \cite{Grozdanov:2019uhi}, which connects to the breakdown of hydrodynamics. Including further dimensionless parameters produces exceptional lines (or more generally, codimension-2 submanifolds) in parameter space \cite{Cavalcante:2026vgr}. See \cite{Berti:2025hly} for a recent review which includes discussion of some of these topics.

Schematic reasoning for this amplification is as follows: Consider a family of black hole spacetimes which depend continuously on two real dimensionless parameters $\alpha$ and $\beta$. In the case we will consider in this paper, the QNM spectrum of these black holes comes from solving
\begin{equation}\label{eq: Wron cond}
    W(\omega, \alpha, \beta) = 0
\end{equation}
for $\omega$, where $W(\omega, \alpha, \beta)$ is the Wronskian of some ODE between a solution which is ingoing at the future event horizon and a solution which obeys the appropriate falloff at an asymptotic boundary (\eg outgoing at future null infinity for asymptotically flat black holes). Let us further consider the case of an exceptional QNM, which, in addition to Eq.~(\ref{eq: Wron cond}), also satisfies
\begin{equation}
    \frac{\partial W}{\partial \omega}(\omega, \alpha, \beta) = 0\, .
\end{equation}
We consider an isolated double root with $\partial_\omega^2W\neq0$ and one independent QNM eigenfunction. This excludes an ordinary crossing of modes in independent sectors. We assume that $W$ is locally holomorphic in $\omega$ and sufficiently smooth in the parameters. Since $W$ is generally complex, the double-root condition imposes two additional real conditions; a nondegenerate two-parameter unfolding therefore gives an isolated EP when one is present.

An exceptional point (EP) in the parameter space of black holes is a pair $(\alpha_E, \beta_E)$ for which there exists an exceptional QNM $\omega_E$. Near this EP, we can Taylor expand the Wronskian as
\begin{equation}
    W(\omega, \alpha, \beta) \approx \eta_1 (\omega - \omega_E)^2 + \eta_2 (\alpha - \alpha_E) + \eta_3(\beta - \beta_E)
\end{equation}
for some constants $\eta_{1,2,3}$.

The QNM condition now picks out two QNMs in the neighbourhood of $\omega_E$, as expected since $W$ has a double root. At leading order, these have frequencies
\begin{equation}\label{eq: GR local splitting}
    \omega_\pm \simeq \omega_E \pm \sqrt{\eta_\alpha(\alpha-\alpha_E)+\eta_\beta(\beta-\beta_E)}\, ,
\end{equation}
where $\eta_\alpha=-\eta_2/\eta_1$ and $\eta_\beta=-\eta_3/\eta_1$. The omitted regular displacement of the mean frequency is subleading to the square-root term. QNM hysteresis is the multivaluedness of this function: following the two QNMs once around the EP exchanges them, and following them twice returns their frequencies to their starting values.

We now consider a small dimensionless coupling $c$. It is useful to retain the regular motion of the mean frequency by defining
\begin{equation}\label{eq: pair invariants}
    S=\frac{\omega_++\omega_-}{2}\,,\qquad
    D=\frac{(\omega_+-\omega_-)^2}{4}\,,\qquad
    \omega_\pm=S\pm\sqrt D\, .
\end{equation}
For an isolated generic double root, $S$ and $D$ are smooth and single-valued even when the individual frequencies exchange labels. Near the uncorrected EP,
\begin{equation}\label{eq: local discriminant}
    D(\alpha,\beta,c)=\eta_\alpha(\alpha-\alpha_E)+\eta_\beta(\beta-\beta_E)+\eta_c\,c
    +o\bigl(|\alpha-\alpha_E|+|\beta-\beta_E|+|c|\bigr)\, .
\end{equation}
The ordinary first-order frequency correction at fixed $\alpha,\beta$ is therefore
\begin{equation}\label{eq: singular frequency derivative}
    \left.\partial_c\omega_\pm\right|_{c=0}
    =\left.\partial_cS\right|_{c=0}
    \pm\frac{\left.\partial_cD\right|_{c=0}}{2\sqrt{D(\alpha,\beta,0)}}\, .
\end{equation}
Provided $\eta_c=\partial_cD|_E\neq0$, we expect an inverse-square-root divergence as $(\alpha,\beta)\to(\alpha_E,\beta_E)$. A special EFT deformation tangent to the exceptional locus can have $\eta_c=0$, so the enhancement is generic rather than inevitable for every choice of operator.

The same local expansion determines the leading behaviour at small finite coupling. Keeping the leading splitting terms gives
\begin{equation}\label{eq: resummed omega c}
    \omega_\pm^c\simeq\omega_E\pm
    \sqrt{\eta_\alpha(\alpha-\alpha_E)+\eta_\beta(\beta-\beta_E)+\eta_c\,c}\, ,
\end{equation}
where now we use a ${}^c$ superscript to denote finite-$c$ coupling values.

The EP at $(\alpha_E,\beta_E)$ is the intersection of the $c=0$ plane with an exceptional line. This line is locally defined by $D=0$, which imposes two real conditions on $\alpha,\beta,c$. Its displacement obeys
\begin{equation}\label{eq: abstract EP motion}
    \eta_c=-\eta_\alpha\,\partial_c\alpha_E^c-\eta_\beta\,\partial_c\beta_E^c\, .
\end{equation}
We assume that the real Jacobian with columns $(\Re \,\eta_\alpha,\Im \,\eta_\alpha)^T$ and $(\Re \, \eta_\beta,\Im \, \eta_\beta)^T$ is nonsingular. At the fixed uncorrected parameters, the leading splitting is
\begin{equation}\label{eq: EP square root}
    \omega_\pm(\alpha_E,\beta_E,c)=\omega_E\pm\sqrt{\eta_c\,c}+o(\sqrt{|c|})\, .
\end{equation}
Following the shifted exceptional line instead keeps the modes degenerate; their common frequency has a regular first-order displacement. These are different comparisons of black holes in the corrected theory.

Away from the EP, expanding the square root in $c$ requires
\begin{equation}\label{eq: spectral expansion validity}
    |c\,\partial_cD|\ll|D_0|\,,\qquad D_0=D(\alpha,\beta,0)\, .
\end{equation}
Near the EP this reduces to $|c|\ll|\omega-\omega_E|^2/|\eta_c|$. A leading approximation across the nearly degenerate region is instead
\begin{equation}\label{eq: uniform pair approximation}
    \omega_\pm\simeq S_0+c\,S_c\pm\sqrt{D_0+c\,D_c}\, ,
\end{equation}
where $S_c=\partial_cS|_{c=0}$ and $D_c=\partial_cD|_{c=0}$ at fixed black hole parameters. This reorganises the frequency expansion while keeping the leading EFT operator. Its accuracy is limited by the omitted terms in the derivative expansion, and it does not give uniform relative accuracy for a gap that vanishes at the shifted EP. In string theory those omitted terms should be counted in $\alpha'$, rather than assuming an all-orders Taylor series in $c\propto\alpha'^3$.

The local spectral assumptions deserve some qualification. It is not clear that a scalar Wronskian condition can always be formulated without sufficient symmetry to reduce the linearised Einstein equation to an ODE (although recent work shows that less symmetric families also allow the determination of QNMs by solving ODEs \cite{Hintz:2026jiw,Santos:2026pbu,Yoshida:2026yly}). Discretising a PDE gives a matrix problem with a smooth characteristic determinant and motivates the same argument. Extending it to the continuum requires the coalescing pair to remain an isolated spectral sector with the stated regularity. Our explicit example satisfies an ODE boundary-value problem, so it avoids this additional issue.

In this paper, we will give a concrete example by studying corrections to the QNM spectrum of equal-spin Myers--Perry-AdS$_5$ (MPAdS$_5$), labelled by two dimensionless combinations of black hole radius, spin and AdS radius. This is a relevant example as MPAdS$_5$ lies in the universal, purely gravitational sector of holographic theories of gravity. Furthermore, these spacetimes are cohomogeneity-one, with symmetry dictating that certain tensor modes decouple from the rest \cite{Kunduri:2006qa,Murata:2007gv,Murata:2008yx}. QNMs are then determined by solving an ODE, simplifying the analysis. We identify an exceptional QNM and demonstrate the enhanced first-order corrections nearby. Treating the pair together, we calculate the splitting coefficient and the displacement of the EP at fixed physical ensemble conventions. For planar $\mathcal N=4$ SYM this gives a leading $\lambda^{-3/4}\sim(\alpha'/L^2)^{3/2}$ splitting at the uncorrected EP, even though the leading string correction to the action and equations of motion is $\mathcal O(\alpha'^3)$ \cite{Gross:1986iv}. The top-down setting fixes the EFT operator and its coupling; the local mechanism applies more generally whenever that operator unfolds an isolated exceptional pair with nonzero $\eta_c$.

An outline of this paper is as follows: In Section~\ref{sec: analysis} we state our approach to EFT-correcting MPAdS$_5$ QNMs. In Section~\ref{sec: results} we present our numerical methods and results for the amplified corrections, the leading finite-coupling splitting and the displacement of the EP. In Section~\ref{sec: discuss} we discuss the interpretation and possible relevance to other areas of black hole physics.

\section{$\lambda^{-3/2}$ corrections to MPAdS$_5$ QNMs}\label{sec: analysis}
\subsection{Corrections to IIB SUGRA}\label{sec: SUGRA}
We will consider the $\alpha'^3$ higher-derivative corrections to IIB SUGRA arising from string theory. It is known that supersymmetry demands that the leading-order corrections to this theory appear at order $\alpha'^3$ \cite{Gross:1986iv,Freeman:1986zh,Grisaru:1986vi}. After a Freund–Rubin compactification on $S^5$ \cite{Freund:1980xh}, the dynamics of the \emph{gravitational} sector on ${\rm AdS}_5$ in the $N \to \infty$ limit are given by extremising the action
\begin{equation}\label{eq: string action}
    S_{\rm IIB}[g] = \frac{1}{16 \pi G_5}\int \dd^5 x\sqrt{-g} \left( R + \frac{12}{L^2} + \gamma \,\mathcal W \right)\, , 
\end{equation}
where $G_5$ is Newton's constant in 5D, $L$ is the AdS radius, 
\begin{equation}\label{eq: gamma def}
    \gamma = \alpha'^3\frac{\zeta(3)}{8} 
\end{equation}
and
\begin{equation}
    \mathcal W = C^{aebf}\, C_{a\phantom{g}b}^{\phantom{a}g\phantom{b}h} \, C^{c\phantom{e}d}_{\phantom{c}e\phantom{d}g}\, C_{cfdh} - \frac{1}{4}C^{abef}\, C_{ab}^{\phantom{ab}gh} \, C^{cd}_{\phantom{cd}eg}\, C_{cdfh} \, .
\end{equation}
One might expect that, since the background self-dual 5-form $F_5$ is non-zero, its higher-derivative couplings and the other ten-dimensional fields must also be included. Representation theoretic cancellations and the five-dimensional reduction justify using the Weyl-quartic action for this gravitational sector at the order considered \cite{Paulos:2008tn,Buchel:2008universal}. Additional fields can be sourced at this order, while their feedback into the relevant five-dimensional Einstein-frame metric equations starts beyond it. The equation of motion derived from this action reads
\begin{equation}\label{eq: GREFT eom}
    R_{ab} - \frac{1}{2} g_{ab} R - \frac{6}{L^2} g_{ab} = \gamma \, T_{ab}^{\rm{eff}}[g] \left(1 + \mathcal O\left(\alpha' K^{2}\right)\right)
\end{equation}
where $K$ is the characteristic curvature inverse-length scale of the metric $g_{ab}$, and $T^{\rm{eff}}_{ab}[g]$ is the effective energy-momentum tensor, listed in Appendix~\ref{app: eff EM tensor}.

The 't Hooft coupling of the boundary field theory is
\begin{equation}
    \lambda = \frac{L^4}{\alpha'^2}\, .
\end{equation}
Relating this action back to the EFT approach discussed in Section~\ref{sec: intro}, we are therefore expanding in the small dimensionless coupling
\begin{equation}
    c := \frac{\zeta(3)}{8} \frac{\alpha'^3}{L^6} = \frac{\zeta(3)}{8} \lambda^{-3/2} \,,
\end{equation}
and so we will be computing $\lambda^{-3/2}$ corrections to boundary observables. The regime of validity for Eq.~(\ref{eq: GREFT eom}) is that
\begin{equation}\label{eq: EFT validity}
K \ll \alpha'^{-1/2} = \lambda^{1/4}/L\, .
\end{equation}

We find solutions to Eq.~(\ref{eq: GREFT eom}) by linearising about a known solution to GR. That is, we expand the metric as
\begin{equation}
    g_{ab} = g_{ab}^{(0)} + c \, g_{ab}^{(c)}\,,
\end{equation}
where $g_{ab}^{(0)}$ solves the Einstein equation, and where $g_{ab}^{(c)}$ is some perturbation which accounts for the EFT correction to this solution. The equation of motion for $g^{(c)}_{ab}$ will now read
\begin{equation}\label{eq: EFT background corr}
    \mathcal L^{L,\,g^{(0)}}_{ab}[g^{(c)}]= L^6\, T^{\rm eff}_{ab}[g^{(0)}]\, ,
\end{equation}
where $\mathcal L^{L,\,g^{(0)}}_{ab}$ is the linearisation of the LHS of the Einstein equation Eq.~(\ref{eq: GREFT eom}) around the background $g^{(0)}_{ab}$. Typically we will be interested in EFT-corrected black holes, \ie $g_{ab}^{(0)}$ will be some AdS black hole metric, and we will seek a solution $g^{(c)}_{ab}$ to Eq.~(\ref{eq: EFT background corr}) which is stationary and has the relevant spatial symmetry.

To study the linearised dynamics of black holes in EFT, it is helpful to instead expand in two parameters
\begin{equation}
    g_{ab} = g_{ab}^{(0)} + b\, g_{ab}^{(b)} + c\, g_{ab}^{(c)} + b\,c\, g_{ab}^{(bc)}
\end{equation}
where now $b\ll 1$ is the perturbatively small amplitude of a linear perturbation to the stationary spacetime. Taking $c = 0$, we see that $g^{(b)}_{ab}$ obeys the linearised Einstein equation
\begin{equation}
    \mathcal L^{L,\,g^{(0)}}_{ab}[g^{(b)}] = 0\,.
\end{equation}
It is convenient to define the `dynamical' part of the metric as
\begin{equation} \label{eq: dyn metric def}
    h_{ab} = g_{ab}^{(b)} + c \, g_{ab}^{(bc)}\,,
\end{equation}
so that now the equation of motion for this perturbation reads
\begin{equation}\label{eq: dyn metric eom}
    \mathcal L^{L,\,g^{(0)} + c \, g^{(c)}}_{ab}[h] = c \, L^6 \,\frac{\partial}{\partial b} \Bigg |_{b=0}T^{\rm eff}_{ab}[g^{(0)} + b g^{(b)}]\, .
\end{equation}
Since our EFT corrections maintain the relevant symmetries of $g^{(0)}$, we can apply the same gauge-fixing and harmonic decomposition to $h$ as we would to solve the linearised Einstein equation in GR. Plugging this gauge-fixed decomposition into Eq.~(\ref{eq: dyn metric eom}) will now give a second-order ODE for metric components which we can solve using standard numerical methods. This is the approach we will take for MPAdS$_5$ QNMs in EFT.
Finally, the derivative expansion must remain valid for the perturbation as well as for the stationary background. Besides the curvature scale $K$, this involves the proper temporal and spatial variation scales of the mode, as discussed in \cite{Reall:2021voz}. These are local bulk scales. A frequency measured in the chosen boundary time coordinate is not by itself a local inverse wavelength, since redshift and rotation must also be taken into account. We keep this derivative-expansion requirement separate from the small-gap condition in Eq.~(\ref{eq: spectral expansion validity}).

\subsection{Corrections to equal-spin Myers–Perry black hole geometries} \label{sec: EFT MP Corrections}
The Myers–Perry solutions describing rotating black holes in higher dimensions \cite{Myers:1986un} can be constructed with a negative cosmological constant \cite{Hawking:1998kw,Gibbons:2004uw}. In AdS$_5$, these solutions are labelled by their mass and by two spin parameters $a_1, a_2$, generically having a U$(1)^2$ spatial symmetry. When the two spin parameters are taken to be equal, this symmetry is enhanced to U$(2)$, which allows us to describe rotating black holes as cohomogeneity-one solutions, \ie with metric components only depending on a radial coordinate $r$. These metrics take the form
\begin{equation}\label{eq: equal spin ansatz}
    \dd s^2 =- A_1(r) \dd t^2 + A_2(r) \dd r^2 + A_3(r) (\dd t - A_4(r) \sigma_3)^2 + A_5(r) (\sigma_1^2 + \sigma_2^2 + \sigma_3^2)
\end{equation}
where $A_i(r)$ are functions to be determined by the equations of motion, and
\begin{equation}
    \begin{aligned}
        \sigma_1&= \cos \psi \,\sqrt{1-x^2}\, \dd \phi + \sin \psi \,\frac{\dd x}{\sqrt{1-x^2}}\\
        \sigma_2&= \sin \psi\, \sqrt{1-x^2} \,\dd \phi - \cos \psi\, \frac{\dd x}{\sqrt{1-x^2}}\\
        \sigma_3&= \dd \psi + x\, \dd \phi
    \end{aligned}
\end{equation}
are 1-forms such that $\sigma_3$ and $\sigma_1^2+\sigma_2^2$ are invariant under ${\rm U}(2) \leq {\rm SO}(4)$ which acts on the internal $S^3$. The coordinates in this metric take the ranges $x \in (-1,1)$, $\phi \in [0,2\pi)$, $\psi \in [0,4\pi)$.

To study rotating black holes of this kind in EFT, we seek solutions perturbatively close to those found in \cite{Gibbons:2004uw}, \ie 
\begin{equation}\label{eq: background functions}
\begin{aligned}
    A_1(r) &= 1-\frac{a^2}{L^2} - \frac{r^2}{L^2} + \frac{(a^2 + r^2)^3}{2 \mu L^4} + c \, \delta A_1(r)\\
    A_2(r) &= \frac{r^2(a^2 + r^2)}{(a^2 + r^2)^2 \left(1+ \frac{r^2}{L^2}\right) -2 \mu r^2} + c \, \delta A_2(r)\\
    A_3(r) &= \frac{1}{2 \mu (a^2+r^2)}\left(2 \mu - \frac{(a^2+r^2)^2}{L^2} \right)^2 + c \, \delta A_3(r)\\
    A_4(r) &= \frac{a \mu}{\left(1 - \frac{a^2}{L^2}  \right)\left(2 \mu - \frac{(a^2+r^2)^2}{L^2} \right)} + c \, \delta A_4(r)\\
    A_5(r) &= \frac{r^2 + a^2}{4 \left(1-\frac{a^2}{L^2}\right)} + c \, \delta A_5(r)\, 
\end{aligned}
\end{equation}
for some functions $\delta A_i$ to be determined. Here we treat $\mu,a$ as bookkeeping parameters describing the mass and spin of the \emph{unperturbed} geometry, not necessarily corresponding to the mass and spin of the EFT-corrected geometry. 

The $c=0$ black hole has event horizon at $r=r_H$, where $r_H$ is the largest positive root of
\begin{equation}
    (a^2+r_H^2)^2\left(1+\frac{r_H^2}{L^2}\right) - 2 \mu r_H^2 = 0\, .
\end{equation}
Furthermore, this metric is only of Lorentzian signature when $|a|<L$. The functions in Eq.~(\ref{eq: background functions}) are a parametrisation of the metric, and need not each be regular. In particular, $A_4^{(0)}$ has an exterior pole where $2\mu-(r^2+a^2)^2/L^2=0$, while the combinations $A_3A_4$ and $A_3A_4^2$ entering the metric remain finite. Regular metric combinations, or an equivalent factorisation, must therefore be used when specifying smooth numerical unknowns.

Note that in the metric Eq.~(\ref{eq: equal spin ansatz}) there is a remaining coordinate freedom in the definition of $r$. We can mostly fix this freedom by setting $\delta A_2(r) = 0$, which leaves one residual coordinate degree of freedom given by redefining $r$ in such a way as to preserve the $g_{rr}$ component of the metric. We fix this residual coordinate freedom by enforcing that the future event horizon has the radial coordinate $r = r_H$ after transforming $t, \psi$ appropriately to continue this metric across the event horizon in a way which we will outline later. Fixing $r$ in this way ensures that the choice of the Fefferman–Graham (FG) coordinate 
\begin{equation}\label{eq: FG coord def}
    z = z_0 \exp \left(-\int^r \frac{\sqrt{A_2(\tilde r)}}{L} \dd \tilde r \right)
\end{equation}
remains valid, where $z_0$ is chosen such that $z \sim L^2 \sqrt{1-a^2/L^2} \,r^{-1}$ as $r \to \infty$. With this FG coordinate, the representative of the conformal class of boundary metrics is given by
\begin{equation}
    \dd s_{\rm bdy}^2 = - \dd t^2 + \frac{a^2}{L^2} \dd t^2 + a \dd t \, \sigma_3 + \frac{L^2}{4}(\sigma_1^2+\sigma_2^2+\sigma_3^2)
\end{equation}
which, after the coordinate change $\psi = \tilde \psi - 2 a\,  t/L^2$, describes the Einstein static universe $\mathbb R \times S^3$.

We EFT correct the metric so as to preserve the conformal class of boundary metrics \emph{and} the representative of that class with respect to the FG coordinate Eq.~(\ref{eq: FG coord def}). We therefore demand that all our corrections $\delta A_i(r)$ fall off with appropriate powers of $L/r$ as $r \to \infty$ with the exponents predicted by the FG expansion in $D=5$, following the analysis of \cite{deHaro:2000vlm}.

To study physics at the horizon, it will be useful to change coordinates. We define
\begin{equation}
\begin{aligned}
    v &= t +\int^r \dd \tilde r \sqrt{\frac{A_6(\tilde r) A_2(\tilde r)}{A_1(\tilde r)A_6(\tilde r) - A_3(\tilde r) A_5(\tilde r)}}\\
    \chi &= \psi + \int^r \dd \tilde r \,A_3(\tilde r) A_4(\tilde r) \sqrt{\frac{A_2(\tilde r) }{A_6(\tilde r)[A_1(\tilde r) A_6(\tilde r) - A_3(\tilde r)A_5(\tilde r)]}}
\end{aligned}
\end{equation}
where
\begin{equation}
    A_6(r) := A_3(r) A_4(r)^2 + A_5(r)\, .
\end{equation}
In this coordinate system, the metric takes the form
\begin{equation}
    \dd s^2 =- A_1(r) \dd v^2 + 2 A_7(r) \dd r \, \dd v + A_3(r) (\dd v - A_4(r) \tilde \sigma_3)^2 + A_5(r) (\sigma_1^2 + \sigma_2^2 + \tilde \sigma_3^2)
\end{equation}
where $\tilde \sigma_3 = \dd \chi + x \,\dd \phi$ and 
\begin{equation}
    A_7(r) := \sqrt{\frac{A_2(r) [A_1(r) A_6(r) - A_3(r) A_5(r)]}{A_6(r)}}\,.
\end{equation}
Since $A_2(r) \sim (r-r_H)^{-1}$ as $r \to r_H$, to keep $r = r_H$ after EFT corrections we must enforce the condition
\begin{equation}
    A_1(r_H) A_6(r_H) - A_3(r_H) A_5(r_H) =0
\end{equation}
which enforces a Killing horizon at $r=r_H$. From this form of the metric, we can also read off the surface gravity $\varkappa$ and angular velocity $\Omega'_\psi$ at the horizon
\begin{equation}
\begin{aligned}
   \Omega_\psi'  &= \frac{A_3(r_H) - A_1(r_H)}{A_3(r_H) A_4(r_H) }\\
   \varkappa &=-\frac{1}{2 A_7(r_H)} \frac{\dd}{\dd r} \left[A_5(r) \Omega_\psi'^2 + A_3(r) (1 - A_4(r) \Omega_\psi')^2 - A_1(r) \right]\Bigg|_{r=r_H}\, ,
\end{aligned}
\end{equation}
recalling that the Hawking temperature is given by $T_H = \varkappa/2 \pi$. We have denoted the angular velocity here with a $'$ superscript since, following the discussion of \cite{Gibbons:2004ai}, this is the angular velocity measured with respect to the $\psi$ coordinate at infinity. However, as pointed out previously, we should actually make a coordinate redefinition at infinity to put the boundary conformal class representative into the form of an Einstein static universe. This transformation means that the angular velocity as measured in this non-rotating frame, \ie the physical quantity relevant to holographic discussions, is given by
\begin{equation}
    \Omega_\psi = \Omega_\psi' + 2 \frac{a}{L^2} \, .
\end{equation}
For calculations involving QNMs, it is useful to note the near-horizon form of the $v$ and $\chi$ coordinates as $r \to r_H$,
\begin{equation}
\begin{aligned}
    v &\sim t + \frac{1}{2\varkappa} \log(r/r_H-1)\\
    \chi &\sim \psi + \frac{\Omega_\psi'}{2\varkappa} \log(r/r_H-1)\, .
\end{aligned}
\end{equation}
Substituting the ansatz Eq.~(\ref{eq: equal spin ansatz}) into the equation of motion Eq.~(\ref{eq: GREFT eom}) gives a system of five equations, \ie the components of the EFT-corrected Einstein equation which are not forced to vanish by symmetry. However, it is sufficient to solve only four of these equations, with the Bianchi identity imposing that the fifth equation vanishes exactly in this case. We are then free to ignore this last equation, which we now denote as $\mathcal C= 0$. We will numerically verify the vanishing of $\mathcal C$ when we solve for the EFT corrections to MPAdS$_5$, providing a strong test of our analytic and numerical calculations.

The four equations to solve are second-order ODEs for $\delta A_{1,3,4}(r)$ and a first-order ODE for $\delta A_5(r)$. Enforcing the decay at infinity as described above gives four boundary conditions. We also enforce that $\Omega_\psi$ is held fixed under these perturbations. Enforcing that the event horizon lies at $r=r_H$ also provides a boundary condition.

We now have one remaining boundary condition to choose, and one degree of freedom in our solution: perturbing the background parameters $(\mu,a)$ so as to keep the $\Omega_\psi$ fixed. It would be natural to choose to keep the Hawking temperature or the energy as measured at infinity fixed, depending on which ensemble we wish to study. However, a schematic calculation shows that in this coordinate system, \ie with our choice of $g_{rr}$, arbitrary deformations of $(\mu, a)$ produce $\sqrt{r-r_H}$ terms in the components of the metric. These non-smooth terms would be inconvenient for later analysis of the corrections to the QNM spectrum, so we simply choose a  boundary condition at the horizon, derived from the equations of motion, which enforces that the functions $A_{1,3,4,5}(r)$ have finite derivatives at $r =r_H$. This choice, which we call the \emph{numerical ensemble}, does not correspond to keeping an obvious physical observable fixed under these perturbations. Instead, this is just a convenient approach for our numerics, and we must remember at the end of all calculations to transform all observables back into the ensemble of physical interest by taking the relevant thermodynamic derivatives. We now have seven boundary conditions for the ODEs, which are generically of Robin type.

To numerically determine $\delta A_i(r)$, we must take care to avoid complications due to the pole in $A_4^{(0)}$. As such, we will instead make the following ansatz for the metric perturbation:
\begin{equation}
    g^{(c)} = \frac{r_H^2}{r^2}\left[B_1(r) \dd t^2 + 2 L B_2(r) \, \dd t \,\sigma_3 + L^2 B_3(r) \sigma_3^2 + L^2 B_4(r) (\sigma_1^2 + \sigma_2^2 + \sigma_3^2)\right]
\end{equation}
and convert all our equations of motion and boundary conditions for $\delta A_i(r)$ into equations of motion and boundary conditions for $B_i(r)$. It is then trivial to express $\delta A_i(r)$ in terms of $B_i(r)$. Note that the prefactor of $\frac{r_H^2}{r^2}$ is included so that $B_i(r)$ remain bounded as $r \to \infty$ is the relevant boundary condition to keep the conformal class of boundary metric fixed. The boundary conditions for $B_i(r)$ at $r \to \infty$ are then those imposed by the equations of motion. We list these and solve the equations of motion to find $B_i(r)$ numerically in Section~\ref{sec: results}.

It is more convenient for calculations to parametrise this family of black holes by horizon coordinate-radius $r_H$ and spin parameter $a$. At $c=0$, the Hawking temperature and angular velocity at the horizon are then given by
\begin{equation}
    T_H \Big |_{c=0}= \frac{2 r_H^4 + L^2 r_H^2 - a^2 L^2}{2 \pi r_H L^2 (a^2 + r_H^2)}\, , \qquad \Omega_\psi \Big |_{c=0}= \frac{2 a (L^2 + r_H^2)}{L^2 (a^2 + r_H^2)}\, .
\end{equation}
To avoid a naked singularity, we must have $T_H \geq 0$. In this paper we will in fact demand something stronger, namely that the black holes we are considering are the dominant phase in the grand canonical ensemble (GCE), \ie that the radius obeys $r_H> L$, since $r_H=L$ is the horizon radius in GR at which the Hawking–Page phase transition occurs, independently of $a$ \cite{Hawking:1982dh,Hawking:1998kw}. Under $\lambda^{-3/2}$ corrections, the location of this first-order phase transition will shift in a way which can be analytically determined following the methods of \cite{Reall:2019sah}. The shifting of this phase transition will not concern us in this paper as we stay well within the phase dominated by MPAdS$_5$. Staying within this regime also guarantees that superradiant instabilities are not present, since we can infer that\footnote{Note our factor of 2 difference from standard conventions. This can be traced back to the fact that $\psi$ has periodicity $4 \pi$.} $|\Omega_\psi| L < 2$ \cite{Hawking:1999dp,Kunduri:2006qa}. The right treatment to correct this bound by EFT is unclear, as this comes from an instability of high-angular-momentum modes which would then lie outside the range of validity of EFT (see Section 4.8 of \cite{Reall:2021voz} for a discussion of this. See also \cite{Buchel:2026ajv} for a similar discussion for black branes). By staying in the $r_H>L$ regime, we stay clear of these kinds of concerns.

\subsection{Corrections to QNMs}
We find EFT corrections to the QNMs following the approaches of \cite{Cardoso:2018ptl,Boyce:2025fpr,Boyce:2026rnn}. For equal-spin Myers–Perry black holes, the enhanced symmetry group U$(2)$ allows for a simple decomposition of the gravitational perturbations into harmonics of this group. Furthermore, as shown in \cite{Murata:2007gv,Murata:2008yx}, a particular component of these perturbations decouples from the others, allowing us to evaluate certain QNM frequencies by simply solving an ODE. It is this decoupled component which we will consider in this paper.

The spatial isometries are generated by the Killing vectors
\begin{equation}
\begin{aligned}
    \xi_x &= - \cos \phi \sqrt{1-x^2} \partial_x - \frac{x \sin \phi}{\sqrt{1-x^2}} \partial_\phi + \frac{\sin \phi}{\sqrt{1-x^2}} \partial_\psi\\
    \xi_y &= \sin \phi \sqrt{1-x^2} \partial_x - \frac{x \cos \phi }{\sqrt{1-x^2}}\partial_\phi + \frac{\cos \phi}{\sqrt{1-x^2}} \partial_\psi\\
    \xi_z &=\partial_\phi\\
    e_3 &= \partial_\psi \, .
\end{aligned}
\end{equation}
The isometry algebra $\mathfrak u(2)$ is then represented on the spacetime by the self-adjoint operators
\begin{equation}
    L_i = \ii \,\mathcal L_{\xi_i},\qquad W_3 = \ii \,\mathcal L_{e_3}\, ,
\end{equation}
which obey the algebra
\begin{equation}
    [L_i, L_j] = \ii \, \varepsilon_{ijk} \, L_k\, ,\qquad [L_i, W_3] = 0\,.
\end{equation}
Representations of this algebra are labelled by the eigenvalues of $L^2 := L_iL_i$ and $W_3$.  The Wigner-$D$ functions $D^J_{K,M}(x,\phi,\psi)$ are defined by
\begin{equation}
    L^2 D^J_{KM} = J(J+1) \,D^J_{KM}\, ,\qquad  L_z D^J_{KM} = M \,D^J_{KM}\, ,\qquad  W_3 D^J_{KM} = K\, D^J_{KM}
\end{equation}
for $|M|, |K| \leq J$.
We define 1-forms $\sigma_\pm = \sigma_1 \mp \ii\, \sigma_2$, with respect to which the decoupled metric perturbation is given by
\begin{equation}\label{eq: MP metric pert}
    h_{ab} \, \dd x^a \dd x^b = e^{-\ii \omega t}Z(r) D^J_{JM}(x,\phi,\psi)\sigma_+ \sigma_+
\end{equation}
for any $M$. This transforms in the representation $(J,J+2)$ of the symmetry algebra with $L_z$ eigenvalue $M$. Note that the operators $L_\pm := L_1 \pm \ii L_2$ which raise and lower $M$ are generated by isometries of this spacetime, so the value of $M$ will not appear in any observables. As such, we set $M=J$ without loss of generality. In this case the relevant Wigner-$D$ function is given by (up to an arbitrary choice of normalisation)
\begin{equation}
    D^J_{JJ}(x,\phi,\psi) = e^{-\ii J (\phi+\psi)}(1+x)^J\, .
\end{equation}
Substituting the metric perturbation Eq.~(\ref{eq: MP metric pert}) into the linearised Einstein equations produces a second-order ODE for $Z(r)$. We proceed to put this equation into a convenient numerical form with a similar approach to \cite{Boyce:2025fpr}. Namely, we factorise
\begin{equation}
    Z(r) = b(r) \, Y(r)
\end{equation}
where the boundary regulator function $b(r)$ is defined as
\begin{equation}
    b(r) := \left(1- \frac{r_H}{r} \right)^{- \ii (\omega + (2 + J ) \Omega_\psi')/2\varkappa} \left(\frac{r}{r_H}\right)^2
\end{equation}
where here we use the EFT-corrected values of $\varkappa$ and $\Omega_\psi'$.

With respect to the compactified coordinate $y:=1-r_H/r$, QNMs are determined by solving an equation of the form 
\begin{equation}
    p(y) Y''(y) + q_\omega(y) Y'(y) + r_\omega(y) Y(y) = c \, \left[ f_\omega(y) Y(y) + g_\omega(y) Y'(y) \right]
\end{equation}
for $Y(y)$, with analyticity imposed at the boundaries $y = 0,1$ which correspond to the horizon and the conformal boundary respectively. Here $p(y), q_\omega(y), r_\omega(y)$ are rational functions of $y$ and, implicitly, $J$, $r_H$, and $a$. Furthermore $q_\omega$ and $r_\omega$ are polynomials in $\omega$ of degrees 1 and 2 respectively. The forcing terms $f_\omega(y), g_\omega(y)$ are also rational in $y, J,r_H,a$, and now have a generally rational dependence on $\omega$. They are also constructed from the background corrections $\delta A_i$ and the corrections to the thermodynamic variables $\varkappa$ and $\Omega_\psi'$. Note in the formula for $b(r)$ that we use the angular velocity $\Omega'_\psi$ with respect to the $\psi$ coordinate. That this is the correct combination to appear in the boundary regulator function can be seen by making an appropriate coordinate transformation to extend the metric Eq.~(\ref{eq: MP metric pert}) across $r=r_H$ and finding that the $(1-r_H/r)$ factor is needed. The $(r/r_H)^2$ factor is included to guarantee that the lowest Frobenius index of the relevant ODE for $Y(y)$ at $y=1$ is zero. This lowest Frobenius index corresponds to perturbing the induced metric on the conformal boundary, which we would like to keep fixed \cite{Michalogiorgakis:2006jc, deHaro:2000vlm}. As such, the boundary condition at $y = 1$ is that $Y(1) =0$.

It is now at this final step that we EFT correct the frequency by expanding
\begin{equation}
\begin{aligned}
    \omega &= \omega_0 + c \, \delta \omega \\
    Y(y) &= Y_0(y) + c \, \delta Y(y)\, .
\end{aligned}
\end{equation}
Here, $\omega_0$ and $Y_0(y)$ are the relevant frequency and solution to GR, satisfying
\begin{equation}\label{eq: QNM eom GR}
    \left[ p(y) \frac{\dd^2}{\dd y^2} + q_{\omega_0}(y) \frac{\dd}{\dd y} + r_{\omega_0}(y) \right]Y_0(y) = 0\, ,
\end{equation}
and now $\delta \omega$ and $\delta Y(y)$ are the corrections obeying
\begin{equation}\label{eq: QNM eom EFT}
\begin{aligned}
    \left[ p(y) \frac{\dd^2}{\dd y^2} + q_{\omega_0}(y) \frac{\dd}{\dd y} + r_{\omega_0}(y) \right] \delta Y(y)  &=  \left[f_{\omega_0}  + g_{\omega_0} \frac{\dd}
    {\dd y}\right] Y_0(y)\\ 
    &\quad- \delta \omega \left[\partial_\omega q_\omega(y)\frac{\dd}{\dd y} +\partial_\omega r_{\omega}(y) \right]\Bigg |_{\omega = \omega_0} Y_0(y)\, .
\end{aligned}
\end{equation}
We explain how to numerically determine $\omega_0$ and $\delta \omega$ from these equations in Section~\ref{sec: results}.

Note that we could have alternatively considered metric perturbations constructed from $D^J_{-J M} \sigma_-^2$ which would also decouple. However, these are related by complex conjugation to the $\sigma_+^2$ perturbations we consider and so can be calculated from this spectrum. There is a further discrete transformation which simplifies our analysis: The pullback of the equal-spin Myers–Perry metric with spin $a$ under the map $x \mapsto -x, \psi \mapsto -\psi$ is an equal spin Myers–Perry metric with spin $-a$. Under this transformation, the sign of the $W_3$ eigenvalue changes, and so combining the action of these two transformations on a mode of the spin $a$ black hole with frequency $\omega$ in the $(J,J+2)$ rep with $L_z$ eigenvalue $M$ gives a correspondence between QNMs: 
\begin{equation}
    (J,M,J+2,a,\omega)
    \overset{\text{conjugation}}{\leftrightarrow}
    (J,-M,-(J+2),a,-\omega^*)
    \overset{\text{pullback}}{\leftrightarrow}
    (J,-M,J+2,-a,-\omega^*)\, .
\end{equation}
We therefore conclude that the QNM spectrum is preserved under $(a,\omega) \mapsto (-a, -  \omega^*)$, and so we can choose to restrict to ${\rm Re}(\omega) >0$ and study both signs of $a$, or we can restrict to $a\geq 0$ and study the whole range of ${\rm Re}(\omega)$. In this paper we choose to do the former.

One final complication is that the QNMs we have computed in this section have a $t$ and $\psi$ dependence of the form $e^{- \ii \omega t - \ii (J+2) \psi}$, however we are interested in the frequency as measured on the conformal boundary, \ie as measured by the coordinates $(t, \tilde \psi = \psi + 2 a t/L^2)$. As such, the frequency as measured on the boundary is
\begin{equation}
    \omega_{(t, \tilde \psi)} = \omega_{(t,\psi)} -2 \frac{a (J+2)}{L^2} \, ,
\end{equation}
where $\omega_{(t, \psi)}$ is the frequency $\omega$ we have considered so far in this section. Throughout the rest of this paper we will now use the symbol $\omega$ to denote $\omega_{(t, \tilde \psi)}$.

\section{Results}\label{sec: results}
\subsection{Numerical methods}\label{sec: num methods}
We solve the equations of the previous section numerically using spectral methods, placing the equations on an $N=25$ Chebyshev grid and working to 15 digits of precision. 

Having removed all gauge freedom and enforced the required boundary conditions for the EFT corrections to the MPAdS$_5$ geometry, we compactify with $y=1-r_H/r$ and solve the linear background problem using the built-in \texttt{Mathematica} function \texttt{LinearSolve}. The boundary conditions obeyed by $B_i(y)$ at $y=1$ are
\begin{equation}
\begin{aligned}
    B_1'(1) = B_2'(1) = B_3'(1) &= 0\\
    B_1(1) - 4 \frac{a}{L} B_2(1) - 4 B_3(1) - 12 B_4(1) + 4 \left( \frac{a}{L}\right)^2[B_3(1) + B_4(1)] &=0\, .
\end{aligned}
\end{equation}
Enforcing a null Killing horizon at $y=0$ imposes
\begin{equation}
     B_1(0) + 2L\,\Omega'_\psi\,B_2(0) + L^2\Omega'^2_\psi\,(B_3(0)+B_4(0))  = 0   \, ,
\end{equation}
and fixing $\Omega_\psi$ to receive no corrections enforces a $y=0$ boundary condition on this horizon
\begin{equation}
     \left(\frac{r_H}{L}\right)^2 B_1(0) + \frac{a}{L} \left(\frac{a}{L} B_1(0) + 2 B_2(0) - 2 \left(\frac{a}{L}\right)^2 B_2(0) \right) = 0 \,.
\end{equation}
The final boundary condition comes from evaluating the equations of motion at the horizon, and involves only up to first order derivatives of $B_i$ evaluated at $y=0$. There is also a fourth equation which we should impose at the horizon of this same form, which is simply the first-order equation of motion for $B_4(y)$ evaluated at $y=0$, and so should not be thought of as a new boundary condition. We include these equations of motion and their boundary conditions in an accompanying \texttt{Mathematica} notebook. 

For the QNMs, having factored out the required behaviour at the horizon $y=0$, we need only impose analyticity here, which is achieved by enforcing the equations of motion. At the conformal boundary $y=1$, we impose the Dirichlet boundary condition $Y(1)=0$. The equation for QNMs in GR, Eq.~(\ref{eq: QNM eom GR}), reduces to a quadratic eigenvalue problem,
\begin{equation}\label{eq: quad eval prob}
    \mathbf P(\omega_0)\mathbf Y_0 = \mathbf 0\, , \qquad \text{ where  } \mathbf P(\omega):=\mathbf M_{0} + \omega\, \mathbf M_{1} + \omega^2\, \mathbf M_{2}
\end{equation}
which we readily solve using relaxation methods. Here the matrices $\mathbf M_2$, $\mathbf M_1$, and $\mathbf M_0$ are the discretisations of the relevant operators accompanying powers of $\omega$, but with their boundary rows at $y=1$ set to
\begin{equation}
(\mathbf M_2)_{1i} = (\mathbf M_1)_{1 i} = 0, \qquad(\mathbf M_0)_{1i} = \delta_{1i}    \, ,
\end{equation}
recalling that the Chebyshev index $i=1$ corresponds to this boundary. We pick out the normalisation of $\mathbf Y_0$ by imposing $\ell^T \,\mathbf Y_0 = 1$, where $\ell$ is a vector with a $1$ at the horizon and $0$ elsewhere, effectively setting $\mathbf Y_0 \big|_{\rm horizon}=1$. It will also be helpful for later considerations to solve the dual problem 
\begin{equation}
   \mathbf P(\omega_0)^T\,\mathbf Y_D = \mathbf 0\,,
\end{equation}
in the same way, with $\mathbf Y_D^T \,\ell =1$.

To solve the problem Eq.~(\ref{eq: quad eval prob}) numerically with relaxation methods, we need to specify an initial guess for $\omega$ and $\mathbf Y_0$. To do this, we start by computing the QNM spectrum at $a=0$ for some horizon radius of choice, \eg $r_H = L$. This full spectrum is computed by converting Eq.~(\ref{eq: quad eval prob}) into a generalised eigenvalue problem
\begin{equation}\label{eq: mat QNM solve}
\begin{pmatrix}
\mathbf M_0 & \mathbf M_1 \\
\mathbf 0   & \mathbf I
\end{pmatrix}
\begin{pmatrix}
\mathbf Y_0 \\
\tilde{\mathbf Y}_0
\end{pmatrix}
= \omega
\begin{pmatrix}
\mathbf 0 & -\mathbf M_2 \\
\mathbf I & \mathbf 0
\end{pmatrix}
\begin{pmatrix}
\mathbf Y_0 \\
\tilde{\mathbf Y}_0
\end{pmatrix}
\end{equation}
for some vector $\tilde {\mathbf Y_0}$, which is enforced to be given by $\tilde {\mathbf Y_0} = \omega \mathbf Y_0$ by this matrix problem. We solve Eq.~(\ref{eq: mat QNM solve}) numerically using the built-in \texttt{Mathematica} function \texttt{Eigenvalues}.

Armed with a choice of QNM for a non-rotating black hole, we now iteratively move in $(r_H,a)$ parameter space by small increments and find the QNM for a new horizon radius and spin by using the QNM of the previous pair $(r_H, a)$ as our initial estimate. Of course, our discussion of EPs in Section~\ref{sec: intro} predicts that there should be a non-trivial monodromy for QNMs as we move around in parameter space. To account for this, for each pair $(r_H, a)$ we make a canonical choice of contour in parameter space to reach this from our starting spherical black hole at $(L, 0)$: First, we move along the $a=0$ axis to $(r_H,0)$, and then we move along the $r_H = {\rm const}$ axis to $(r_H, a)$. This means that, upon extending a given QNM from $(L,0)$ to the full parameter space, we should see branch cuts starting from EPs and extending away from $a=0$ at constant $r_H$. Such branch cuts allow us to identify candidate EPs. The predicted $\mathbb Z_2$ frequency monodromy is tested by an exchange after one traversal of a small closed path and a return after two traversals. Our Newton's method solver stops when iterations produce a fractional change less than $10^{-10}$ in $\mathbf Y$.

Discretising the higher-derivative equation Eq.~(\ref{eq: QNM eom EFT}) gives a linear algebra problem of the form
\begin{equation}
    \mathbf P(\omega_0)\, \delta\mathbf Y = -\delta \omega\,  \mathbf P'(\omega_0)\mathbf Y_0 + \mathbf f\, ,
\end{equation}
for some forcing vector $\mathbf f$ constructed from $\mathbf Y_0$, its (discretised) derivative, and the EFT corrections to the metric.

A convenient trick to solve for $\delta \omega $ is to contract this whole expression with $\mathbf Y_D$, the solution to the dual problem. The LHS of this equation will then evaluate to $0$, and so we find
\begin{equation}\label{eq: num delta omega}
    \delta \omega = \frac{\mathbf Y_D^T \,\mathbf f}{\mathbf Y_D^T\, \mathbf P'(\omega_0)\, \mathbf Y_0}\, .
\end{equation}
This allows a fast determination of $\delta \omega$ \emph{without} having to calculate $\delta \mathbf Y$.

The frequency shift in Eq.~(\ref{eq: num delta omega}) is singular when its denominator vanishes, provided the numerator remains nonzero. Appendix~\ref{app: conv} describes this diagnostic and plots its magnitude near the EP.

To convert from the numerical ensemble (as defined in Section~\ref{sec: analysis}) to an ensemble of our choosing, which in our case will be the GCE, we first read off the perturbations to $\varkappa$ from the EFT corrections and note that $\Omega_\psi$ is unperturbed. We then invert the relevant Jacobian to find the right $\widetilde \delta r_H$ and $\widetilde \delta a$ we need to perturb $r_H$ and $a$ by to put this geometry into the GCE. We then compute $\partial_{r_H} \omega_0$ and $\partial_a \omega_0$ numerically using the method described above, and calculate the QNM correction in the GCE via
\begin{equation}\label{eq: ensemble frequency conversion}
    \delta_{\rm GCE} \omega = \delta_{\rm num} \omega + \widetilde \delta r_H \, \partial_{r_H} \omega_0 + \widetilde \delta a \, \partial_{a} \omega_0
\end{equation}
where $\delta_{\rm num} \omega$ is the QNM correction in the numerical ensemble, as calculated in Eq.~(\ref{eq: num delta omega}). Alternatively, we can simply replace $\mathbf f$ in Eq.~(\ref{eq: num delta omega}) by
\begin{equation}\label{eq: forcing ensemble conversion}
    \mathbf f_{\rm GCE}=\mathbf f_{\rm num}
    -\left[\widetilde\delta r_H\,\partial_{r_H}\mathbf P(\omega_0)+
    \widetilde\delta a\,\partial_a \mathbf P(\omega_0)\right]\mathbf Y_0
\end{equation}
with the parameter derivatives taken at fixed boundary frequency. This automatically computes $\delta \omega_{\rm GCE}$. We follow this approach, so that from now on when we write $\mathbf f$ we will always be referring to the forcing term in the GCE, and likewise all other quantities will all be calculated in the GCE.

We now determine the leading finite-coupling splitting directly at the EP. Keeping the terms needed to determine the square-root coefficient, we write
\begin{equation}
\begin{aligned}
\omega_\pm &= \omega_E \pm \sqrt{\eta_c\,c}+c\,\omega_1+\cdots\\
    \mathbf Y_\pm &= \mathbf Y_0\pm\sqrt{\eta_c\,c}\,\mathbf Y_1+c\,\mathbf Y_2+\cdots\, .
\end{aligned}
\end{equation}
Now solving order by order, we see that the $\mathcal O(1)$ terms
\begin{equation}
    \mathbf P(\omega_E) \, \mathbf Y_0 = 0
\end{equation}
demand that $\omega_E$ is a QNM. Working to $\mathcal O(\sqrt{c})$,
\begin{equation}\label{eq: Y1 def}
\mathbf P(\omega_E) \, \mathbf Y_1 +\mathbf P'(\omega_E) \, \mathbf Y_0 = 0
\end{equation}
defines a generalised eigenvector $\mathbf Y_1$. Together with $\mathbf Y_0$, it forms a Jordan chain of the quadratic pencil, or equivalently of its companion linearisation. Eq.~(\ref{eq: Y1 def}) allows us to use Newton's method to approach an EP by solving the two equations above while varying $(r_H,a)$, with the QNM boundary conditions also imposed on $\mathbf Y_1$. Two additional normalisation conditions remove the common rescaling and the freedom to add $\mathbf Y_0$ to $\mathbf Y_1$. We impose $\ell^T\mathbf Y_0=1$ as before, and set $\ell^T\mathbf Y_1=0$ to uniquely determine this vector.

Finally, the $\mathcal O(c)$ equation gives
\begin{equation}
    \mathbf P(\omega_E)\, \mathbf Y_2+\omega_1 \mathbf P'(\omega_E) \, \mathbf Y_0
    +\eta_c \,\frac{1}{2}\mathbf P''(\omega_E) \, \mathbf Y_0 + \eta_c \, \mathbf P'(\omega_E) \mathbf Y_1 =\mathbf f \, .
\end{equation}
Applying $\mathbf Y_D^T$ to the left of this equation and noting that, by virtue of Eq.~(\ref{eq: Y1 def}), $\mathbf Y_D^T\, \mathbf P'(\omega_E) \, \mathbf Y_0$ also vanishes, we can then extract
\begin{equation}\label{eq: wc extraction}
    \eta_c = \frac{\mathbf Y_D^T \, \mathbf f}{\mathbf Y_D^T \left[\mathbf P'(\omega_E)\mathbf Y_1 + \frac{1}{2}\mathbf P''(\omega_E) \mathbf Y_0 \right]}\, .
\end{equation}
We verify numerically that the denominator here is non-zero, as expected for a generic second-order EP. The ratio is unchanged by the allowed normalisation changes of the Jordan chain. The displayed expansion organises the retained operator; only its leading square-root coefficient is needed here, and the full string-theory remainder is controlled by the higher orders in $\alpha'$.

Similarly, we can calculate
\begin{equation}
\begin{aligned}
    \eta_{r_H} &= -\frac{\mathbf Y_D^T \, \partial_{r_H}\mathbf P(\omega_E)\,\mathbf Y_0}{\mathbf Y_D^T \left[\mathbf P'(\omega_E)\mathbf Y_1 + \frac{1}{2}\mathbf P''(\omega_E) \mathbf Y_0 \right]}\\
    \eta_a &= -\frac{\mathbf Y_D^T \,\partial_{a}\mathbf P(\omega_E) \mathbf Y_0}{\mathbf Y_D^T \left[\mathbf P'(\omega_E)\mathbf Y_1 + \frac{1}{2}\mathbf P''(\omega_E) \mathbf Y_0 \right]}
\end{aligned}
\end{equation}
from which we can calculate $\eta_\varkappa, \eta_{\Omega_\psi}$ using the Jacobian $\frac{\partial(\varkappa, \Omega_\psi)}{\partial(r_H,a)}$ to replace the $\partial_{r_H}$ and $\partial_a$ in these formulae by the $\partial_{\varkappa}$ and $\partial_{\Omega_\psi}$. We can then calculate the motion of the EP under EFT by
\begin{equation}\label{eq: numerical EP motion}
    \eta_c = -\eta_{\varkappa} \,\partial_c (\varkappa)_E - \eta_{\Omega_\psi} \, \partial_c (\Omega_\psi)_E \, .
\end{equation}

This is one complex equation for two real unknowns and has a unique solution provided
\begin{equation}\label{eq: unfolding determinant}
    \det\begin{pmatrix}\Re \,\eta_{\varkappa}&\Re \,\eta_{\Omega_\psi}\\ \Im \,\eta_{\varkappa}&\Im \,\eta_{\Omega_\psi}\end{pmatrix}\neq0\, \,
\end{equation}
which we verify numerically.

\subsection{Amplified corrections and splitting at an exceptional QNM}\label{sec: num results}
\begin{figure}[t]
    \centering
    \includegraphics[width=\textwidth]{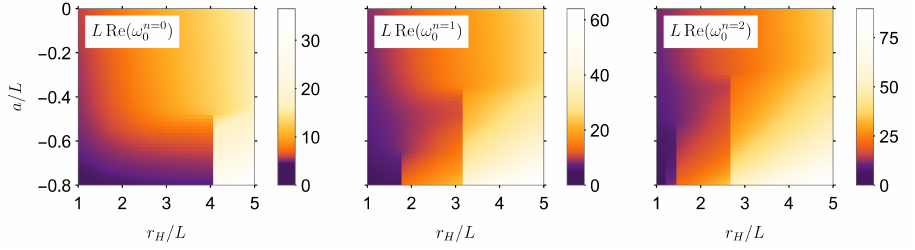}
    \caption{Extension of the $J=0$ tensor fundamental mode and first two overtones of Schwarzschild-AdS$_5$ to QNMs of MPAdS$_5$, following the branching procedure of Section~\ref{sec: num methods}. We see one branch point for $n=0$, two for $n=1$, and three for $n=2$.}
    \label{fig: branching}
\end{figure}

We first identify some EPs and the accompanying exceptional QNMs by following the procedure of Section~\ref{sec: num methods} to calculate QNMs of MPAdS$_{5}$ in the $J=0$ sector. We continue the tensor fundamental mode and first two overtones of Schwarzschild-AdS$_5$ to $a<0$ in Figure~\ref{fig: branching}, where the EPs appear clearly as branch points, accompanied by a branch cut running parallel to the $a/L$ axis (as per our continuation convention). Curiously, for these three overtones in the parameter range $1<r_H/L<5$, $0<a/L<0.9$ we have not found any branch cuts. Under the correspondence $(a, \omega) \mapsto (-a, - \omega^*)$, this corresponds to only finding ${\rm Re}(\omega) <0$ exceptional QNMs in this $a>0$ region.

\begin{figure}[t]
    \centering
    \includegraphics[width=\textwidth]{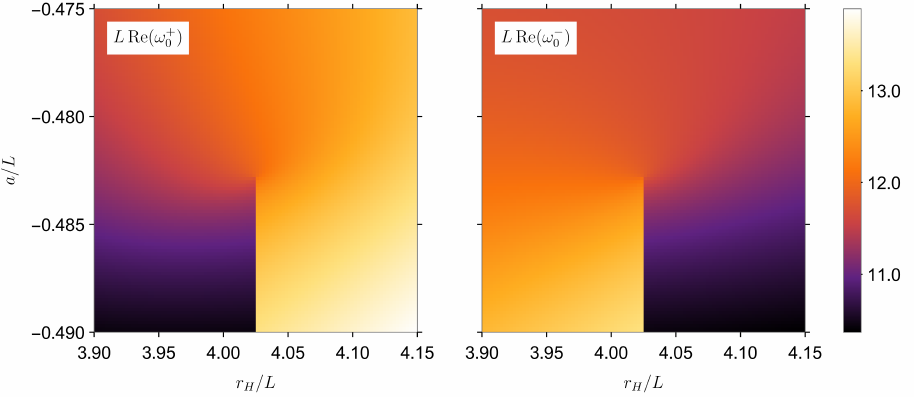}
    \caption{The $n=0$ QNM in Figure~\ref{fig: branching}, continued about the EP.}
    \label{fig: exceptional QNMs}
\end{figure}

\begin{figure}[t]
    \centering
    \includegraphics[width=\textwidth]{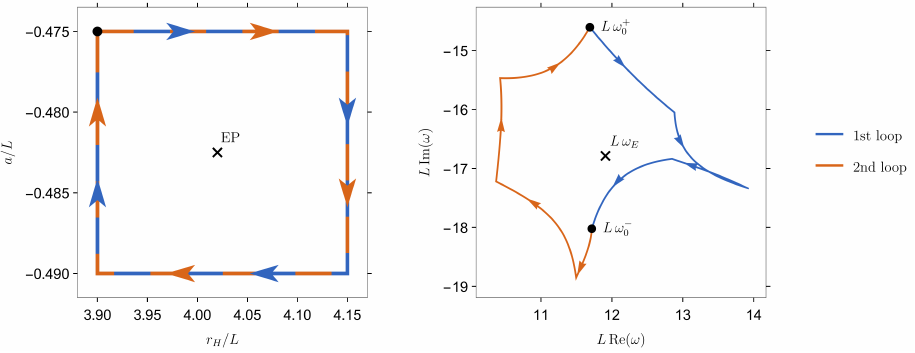}
    \caption{The $\mathbb Z_2$ monodromy about the exceptional point in Figure~\ref{fig: exceptional QNMs}. We follow a closed square-loop twice around the EP, which results in a single trajectory in the $L \, \omega$ plane. The corners of the square appear as kinks in the right hand plot.}
    \label{fig: monodromy}
\end{figure}

\begin{figure}[t]
    \centering
    \includegraphics[width=0.6\textwidth]{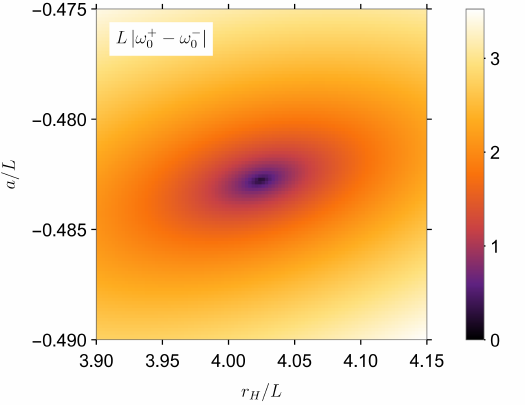}
    \caption{The separation of the QNMs of Figure~\ref{fig: exceptional QNMs}, which tends to zero at the EP.}
    \label{fig: omega sep}
\end{figure}

Figure~\ref{fig: exceptional QNMs} zooms in on the EP reached by continuing the $n=0$ mode shown in Figure~\ref{fig: branching}, using a denser grid. We follow the QNM through the branch cut to identify the QNM which this one becomes degenerate with, and plot both their real parts on this graph, labelling these QNMs by $\omega^{+}_0$ and $\omega^{-}_0$. The square-root structure of the branch point is evident from the monodromy of Figure~\ref{fig: exceptional QNMs}, which we plot explicitly in the complex plane in Figure~\ref{fig: monodromy}. Attempting to follow the $\omega^-$ mode back to Schwarzschild-AdS$_5$ results in the mode moving deep into the complex plane beyond the reach of our numerical methods, having attempted to follow this at $N=100$ and working to 30 digits of precision, so it is unclear exactly which overtone this corresponds to. 

The absolute value of the separation of these $\omega^\pm_0$ is plotted in Figure~\ref{fig: omega sep}. By numerically minimising an interpolation of $|\omega^+ - \omega^-|$ from our data, we identify the location and frequency of this EP as
\begin{equation}\label{eq: EP location}
\begin{aligned}
    (r_H/L)_E &= 4.02\\
    (a/L)_E &=-0.483\\
    (L \, \varkappa)_E &=8.18\\
    (L \, \Omega_\psi)_E &=-1.01\\
    (L \, \omega_0)_E &=11.9-16.8\,\ii
\end{aligned}
\end{equation}
These values agree with those obtained by the augmented Newton solve described above.

\begin{figure}[t]
    \centering
    \includegraphics[width=\textwidth]{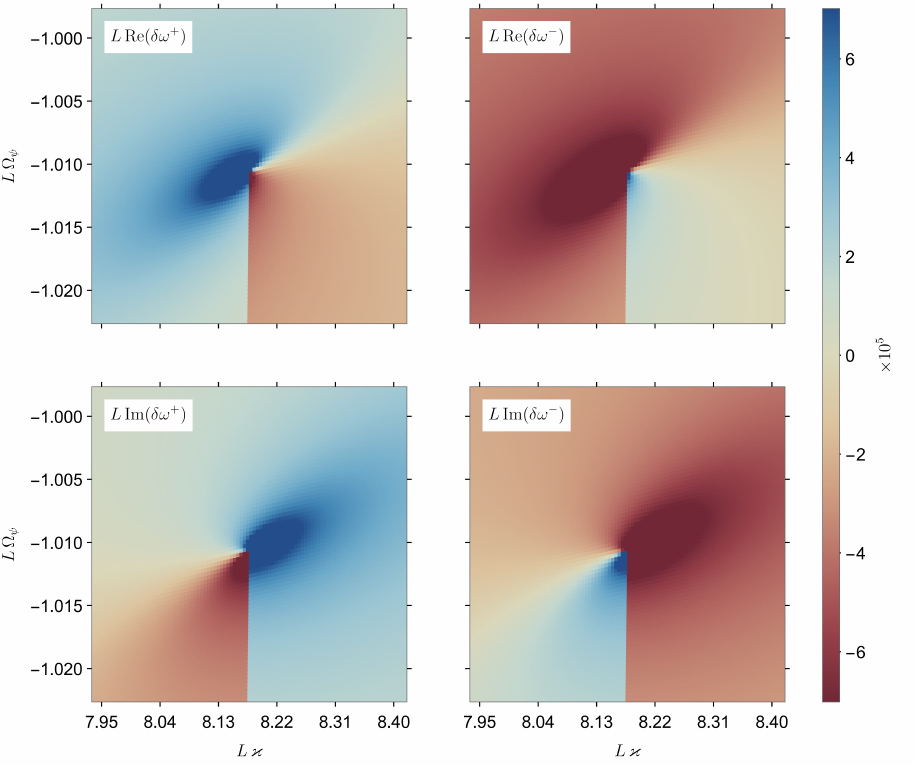}
    \caption{Real and imaginary parts of the coefficients $L\,\delta\omega^\pm$ for the two $J=0$ QNMs near the EP of Figure~\ref{fig: exceptional QNMs}. The actual first-order shifts are $c\,\delta\omega^\pm$. The branch cuts reflect the continuation convention for assigning the individual mode labels.}
    \label{fig: corrections}
\end{figure}

After EFT-correcting these two modes, the large corrections near the EP are displayed in Figure~\ref{fig: corrections}. In all plots of the corrections, we instead parametrise the space of black holes by $(L \, \varkappa, L \, \Omega_\psi)$, which are the appropriate variables to use when working in the GCE. Accordingly, the branch cut is no longer exactly a straight line under this new parametrisation, and is also graphically distorted by the uneven grid spacing after this transformation.

Following the method outlined in Section~\ref{sec: num methods}, we determine the splitting coefficient and the corresponding motion of the EP in the physical $(\varkappa,\Omega_\psi)$ plane:
\begin{equation}\label{eq: splitting and displacement}
\begin{aligned}
    L^2\eta_c&=(-3.84+8.78\,\ii)\times10^5\,,\\
    \partial_c(L\varkappa)_E&=-1.40\times10^5\,,\\
    \partial_c(L\Omega_\psi)_E&=-7.95\times10^3\, .
\end{aligned}
\end{equation}
At the fixed uncorrected GCE parameters of Eq.~(\ref{eq: EP location}), our (unrounded) data predicts the leading finite-coupling behaviour to be
\begin{equation}\label{eq: numerical lambda splitting}
    L\omega^\pm(\lambda)
    =L\omega_E\pm(208+318\,\ii)\lambda^{-3/4}
    +o(\lambda^{-3/4})\, .
\end{equation}
Here we used $c=\zeta(3)\lambda^{-3/2}/8$ and rounded the resulting coefficient to three significant figures. The two signs label the two branches. The exceptional point itself moves according to
\begin{equation}\label{eq: numerical lambda displacement}
\begin{aligned}
    (L\varkappa)_E(\lambda)&=(L\varkappa)_E(\infty)
    -2.11\times10^4\lambda^{-3/2}+o(\lambda^{-3/2})\,,\\
    (L\Omega_\psi)_E(\lambda)&=(L\Omega_\psi)_E(\infty)
    -1.19\times10^3\lambda^{-3/2}+o(\lambda^{-3/2})\, ,
\end{aligned}
\end{equation}
recalling that $\omega_E, \varkappa_E, (\Omega_\psi)_E$ are listed in Eq.~(\ref{eq: EP location}). The square-root splitting compares modes at fixed uncorrected temperature and angular velocity. Following the displaced EP instead keeps the pair degenerate.

\begin{figure}[t]
    \centering
    \includegraphics[width=\textwidth]{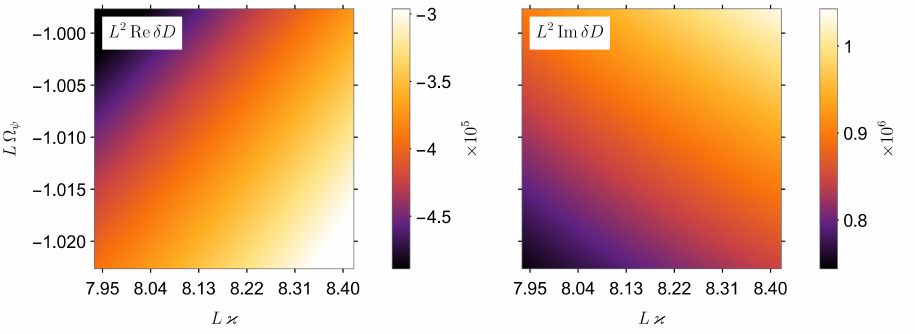}
    \caption{Real and imaginary parts of $L^2\, \delta D$ for the $J=0$ pair. This combination is independent of the exchange of the two mode labels and tends to $L^2 \eta_c$ at the EP, as in Eq.~(\ref{eq: pair limit wc}).}
    \label{fig: test plot}
\end{figure}

We can independently extract $\eta_c$ from the ordinary first-order corrections away from the EP. From Eq.~(\ref{eq: pair invariants}),
\begin{equation}\label{eq: pair limit wc}
    \delta D =(\omega_0^+-\omega_0^-)
    (\delta\omega^+-\delta\omega^-)/2\longrightarrow \eta_c\, .
\end{equation}
Figure~\ref{fig: test plot} shows both components of this continuous, single-valued function. Its non-zero limit provides a check of the predicted reciprocal-gap divergence. 

\begin{figure}[t]
    \centering
    \includegraphics[width=\textwidth]{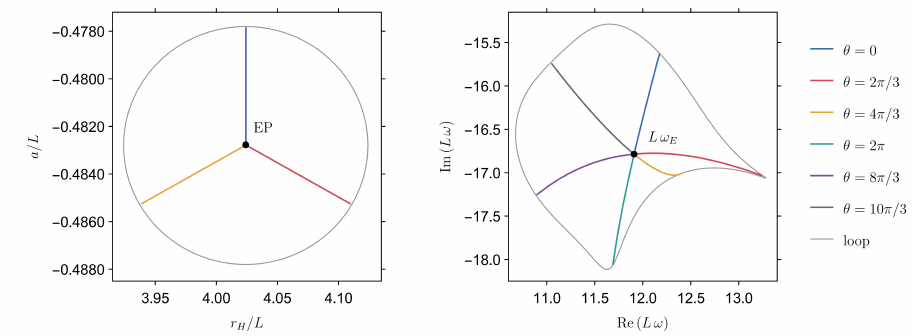}
    \caption{Approach to the exceptional point of Figure~\ref{fig: exceptional QNMs}, and the corresponding approaches of $\omega$. To account for the branch point, we allow for $\theta \in [0,4\pi)$.}
    \label{fig: EP approach}
\end{figure}

\begin{figure}[t]
    \centering
    \includegraphics[width=\textwidth]{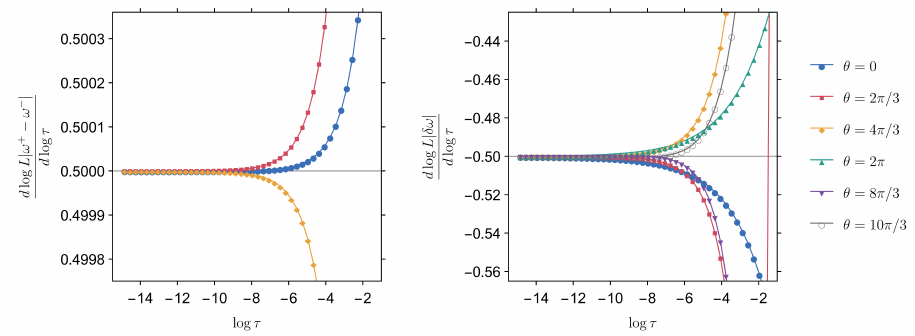}
    \caption{Verification of the $1/2$ and $-1/2$ exponents in $L\,|\omega^+ - \omega^-|$ and $L \,|\delta \omega|$ respectively as we approach the exceptional point as shown in Figure~\ref{fig: EP approach}. We work to $15$ digits of precision with an $N=25$ Chebyshev grid, approaching the exceptional point in Eq.~(\ref{eq: EP location}) (using the unrounded value). On the left hand plot, $\theta$ and $\theta + 2 \pi$ give the same value of $|\omega^+ - \omega^-|$, but specify different $\delta \omega$s in the right hand plot due to the $\mathbb Z_2$ monodromy.}
    \label{fig: slopes}
\end{figure}

\begin{figure}[t]
    \centering
    \includegraphics[width=\textwidth]{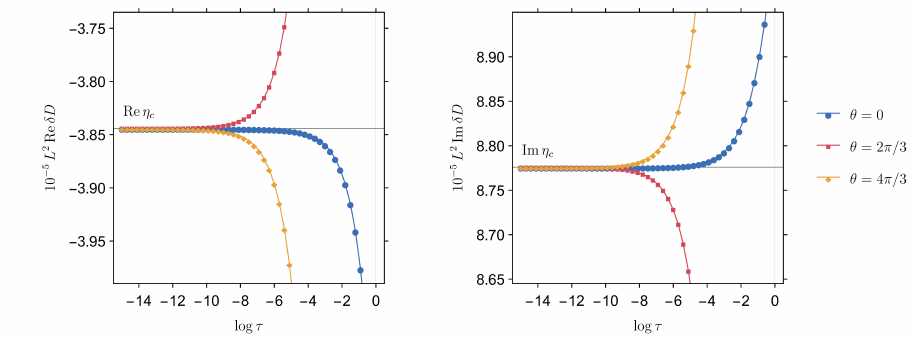}
    \caption{Verification of the numerical value of $\eta_c$ in Eq.~(\ref{eq: splitting and displacement}) as we approach the exceptional point as shown in Figure~\ref{fig: EP approach}.}
    \label{fig: D approach}
\end{figure}

To further verify the expected square-root behaviour near the exceptional point, we also consider some approach trajectories to the exceptional point of the form
\begin{equation}
\begin{aligned}
    r_H &= (r_H)_E +  \tau\, \hat r_H\sin \theta \\ 
    a &= (a)_E +  \tau\, \hat a\cos \theta  
\end{aligned}
\end{equation}
for some fixed angle $\theta$, and constants $\hat r_H, \hat a$ which determine the semi-axes of an ellipse in the $(r_H,a)$ plane. We vary $\tau \to 0$, plotting the approaches of $\omega \to \omega_E$ in Figure~\ref{fig: EP approach}, with $\hat r_H/L = 0.1$ and $\hat a/L = 0.005$. Approaching the EP in this way, we expect
\begin{equation}\label{eq: EP approach behaviour}
\begin{aligned}
    |\omega^+ - \omega^-| &\sim \tau^{1/2}\\
    \delta \omega^\pm &\sim \tau^{-1/2}\, .
\end{aligned}
\end{equation}
To verify this, we would typically plot the logarithms of the quantities in Eq.~(\ref{eq: EP approach behaviour}) divided by $\log\tau$ and note convergence to $1/2$ and $-1/2$ respectively as $\tau \to 0$. However, in our case we expect $L \,\delta \omega \sim 10^5 \tau^{-1/2}$ which would leave a slow-decaying $\sim 10/\log(\tau)$ term which disrupts our ability to read off the limiting value of this plot. As such, we instead plot the nearest-neighbour discretised gradient of these (natural) logarithms against $\log \tau$ from $0$ to $-15$, sampling uniformly. These plots are displayed in Figure~\ref{fig: slopes} where the $\tau^{1/2}$ and $\tau^{-1/2}$ behaviours are clearly visible. We further verify the $\eta_c$ value quoted in Eq.~(\ref{eq: splitting and displacement}) by evaluating $\delta D$ along these approach trajectories in Figure~\ref{fig: D approach}. The limiting values agree with the direct Jordan-chain extraction over the range $-15<\log \tau < -12$ with fractional error at most order $10^{-4}$.

We should also expect from our discussion in Section~\ref{sec: intro} that the functions $Y$ of the $+$ and $-$ branches should coincide at the exceptional point. We test this along the approach trajectories of Figure~\ref{fig: EP approach}, comparing $Y$ between the approaches at $\theta$ and $\theta + 2 \pi$ which correspond to these two branches. We numerically find an order $10^{-5}$ agreement at $\tau = e^{-15}$, supporting this behaviour.

A natural concern about Figure~\ref{fig: corrections} is that, away from the EP, the coefficients can already be of order $L\,\delta\omega\sim10^5$. The small quantity entering the frequency is $c\,\delta\omega$, so the coefficient alone does not establish a loss of perturbative control. Large finite-coupling QNM coefficients also occur away from EPs \cite{Waeber:2015finite}. An ordinary frequency expansion requires a small relative shift, while resolving the two members of a nearly degenerate pair requires the stronger spectral condition in Eq.~(\ref{eq: spectral expansion validity}). A comparison in thermal units is useful for distinguishing this baseline sensitivity from the additional reciprocal-gap enhancement.

These spectral conditions are separate from the bulk derivative expansion. The boundary frequency $\omega$ does not by itself determine the local scale $K$ in Eq.~(\ref{eq: EFT validity}). For example, for Schwarzschild-AdS$_5$ with $f(r)=1+r^2/L^2-m/r^2$, one has $m/r_H^4=L^{-2}+r_H^{-2}$. The horizon curvature remains of order $L^{-2}$ as $r_H/L$ grows, while $\varkappa$ and fixed-overtone boundary frequencies grow as $r_H/L^2$. A local validity estimate must therefore account for the bulk mode profile and its proper variation scales. Our near-EP mechanism concerns the failure of the nondegenerate frequency expansion even when the local derivative hierarchy is maintained. Of course, one should also independently check that the local derivative hierarchy is maintained near these EPs to appropriately describe the background in EFT.

Appendix~\ref{app: conv} shows rapid resolution convergence of $\omega_0^+$ and $\delta\omega^+$ on the sampled parameter grid, with successive-resolution fractional differences below $10^{-8}$ at $N=25$. This sampled test is distinct from direct convergence of the EP location, splitting coefficient and displacement, and from control arbitrarily close to a vanishing gap.

\section{Discussion}\label{sec: discuss}
Our numerical results display the amplification of EFT corrections to individual QNM frequencies near an EP. The ordinary nondegenerate expansion fails as the gap closes, while the Jordan-chain calculation determines a finite leading splitting coefficient. The resulting $\lambda^{-3/4}$ splitting and $\lambda^{-3/2}$ displacement of the EP are predictions of the leading string correction under the local spectral assumptions stated in Section~\ref{sec: intro}. They do not require solving the UV-complete theory at arbitrary coupling.

The frequencies are highly sensitive to black hole and theory parameters near the exceptional line. Following the approaches of \cite{Horowitz:2026fzc,Lam:2026grz}, a useful independent test would be to work with a known theory with second-order equations involving a massive field, determine its near-exceptional spectrum, and then integrate out the massive field. Comparing the full and effective descriptions would test the range of the reorganised expansion and the size of its omitted terms.

A natural question is whether the enhanced pole sensitivity produces a correspondingly large change in the full evolution. The individual pole locations do not settle this question. For example, a local two-pole contribution with denominator $(z-S)^2-D$ has an inverse-transform term proportional to
\begin{equation}\label{eq: regular pair response}
    e^{-\ii S t}\frac{\sin(\sqrt D\,t)}{\sqrt D}
    \longrightarrow t\,e^{-\ii S t}\qquad(D\to0)\, .
\end{equation}
The singular residues of the separate poles cancel in this combination. At the uncorrected EP, $D=\eta_c c+o(c)$, so the displayed pair contribution can have a regular coupling expansion at fixed finite time despite the square-root motion of the individual frequencies. The expansion need not be uniform on times $t\sim|\eta_cc|^{-1/2}$; for a damped pair, these times may lie well beyond its damping time. More general numerators, source overlaps and other spectral contributions are needed for a complete waveform. Related black hole analyses find regular combined responses at EPs and perturbative waveforms despite large spectral rearrangements \cite{Morikawa:2026riesz,OuldElHadj:2026exceptional}.

It remains useful to evolve the gravitational perturbations in the well-posed `fixing-of-equations' approach of \cite{Figueras:2024bba} and compare with the pole calculation. Such an evolution would determine the source-dependent response of this particular system and test how the two-mode description joins the rest of the spectrum. We leave that numerical calculation to future work.

We now turn to the possible relevance to gravitational wave astronomy. The stationary vacuum black holes relevant to astrophysics are described by the Kerr family \cite{Kerr:1963ud,Carter:1971zc,Robinson:1975bv}, parametrised by mass $\mu$ and spin $a$. There is only one dimensionless parameter, $a/\mu$, so isolated second-order EPs on the real Kerr spin axis are not generically expected by parameter counting. This does not exclude special degeneracies enforced by additional structure.

However, as recently detailed in \cite{Kubota:2026hdv}, the Kerr QNM spectrum has other non-analytic features associated with modes crossing the branch cut along the negative imaginary axis. Similar failures of ordinary frequency perturbation theory may occur there.

The Kerr QNM spectrum also displays avoided crossings \cite{Onozawa:1996ux,Motohashi:2024fwt}, where two QNMs pass near each other as $a/\mu$ is varied. A natural interpretation is that nearby EPs occur at complex spin, although a full study for the relevant Kerr branches is still required. A generic EFT perturbation would then produce large but finite corrections along the real-spin axis. Complex-spin EPs can limit the radius of convergence of the spin expansions used in \cite{Cano:2023tmv,Cano:2023jbk,Cano:2024ezp}. A real avoided crossing can nevertheless lie inside that convergence disk, so this observation alone does not rule out a convergent spin expansion. Finite-order truncations can become inefficient near a nearby branch point; an all-spin calculation such as \cite{Boyce:2026rnn}, or a suitable analytic continuation or resummation, would be useful for resolving the enhancement. The all-spin calculation has so far been applied to the fundamental Kerr mode.

The LIGO--Virgo--KAGRA Collaboration has recorded a vast collection of ringdown signals from black hole and neutron star mergers \cite{LIGOScientific:2026sit}. A recent event has allowed the approximate identification of at least two Kerr QNMs \cite{LIGOScientific:2025wao}. The avoided crossings of interest occur for more strongly damped modes, making their measurement challenging. Moreover, the same small gap can amplify other perturbations of the spectral problem, including changes in spin and environmental effects, whenever those perturbations unfold the pair. Large individual excitation factors also need not give a large combined waveform because of the cancellation illustrated in Eq.~(\ref{eq: regular pair response}). Assessing the usefulness of these modes for constraining EFTs therefore requires a calculation of the excited response and its parameter dependence, in addition to the pole shifts.

A common feature of several examples discussed above is that EFT corrections move a locus at which the uncorrected observable is non-analytic. In \cite{Cano:2025ejw}, the amplified corrections arise from shifting the phase boundary in the QNM spectrum labelled by $(\ell,m)$. In the eikonal limit, this boundary is characterised by a critical value of $m/(\ell+1/2)$, separating a region with both damped and zero-damping modes from one with only zero-damping modes. The non-analytic dependence of the frequencies on the distance to this boundary makes the relative corrections to the damping rates large near it. Ref.~\cite{Cano:2025ejw} also distinguishes this failure of the frequency expansion from a failure of the EFT itself.

Similarly, the enhanced near-extremal Kerr QNM corrections in the microcanonical ensemble \cite{Boyce:2026rnn} can be understood in terms of the EFT shift of the extremality bound: holding the energy and angular momentum fixed need not hold the distance to the corrected extremal locus fixed. In our example, the shifted locus is instead the exceptional line, and the enhancement occurs away from extremality. These are distinct spectral problems, but they share the failure of an ordinary expansion about an uncorrected non-analytic locus. When the local derivative hierarchy remains controlled, organising the calculation around the shifted locus can give a useful description without implying a breakdown of EFT.

Finally, it is natural to ask whether other non-analytic black hole observables display similar enhanced corrections. One example is the second-order transition associated with the onset of holographic superconductivity \cite{Hartnoll:2008vx,Hartnoll:2008kx}, where $\langle\mathcal O\rangle\sim\sqrt{T_c-T}$ as $T$ approaches $T_c$ from below. A shift of $T_c$ produces a first-order condensate correction proportional to $(T_c-T)^{-1/2}$ at fixed $T$. As for the QNM pair, this signals a nonuniform expansion about the uncorrected critical point and motivates an expansion about the shifted one. We leave a full study of that example to future work.

\section*{Acknowledgements}
WLB was supported by an STFC studentship ST/Y509127/1. The work of JES was partially supported by STFC consolidated grant ST/X000664/1 and by Hughes Hall College. We used ChatGPT and Codex (OpenAI) as well as Claude (Anthropic) to assist with a critical review of the manuscript and a few aspects of the exceptional-point analysis. The authors take full responsibility for the scientific content and conclusions.

\appendix

\section{The effective energy-momentum tensor for $\alpha'^3$ corrections to IIB SUGRA} \label{app: eff EM tensor}
The effective energy-momentum tensor appearing in Eq.~(\ref{eq: GREFT eom}) is found by varying the action Eq.~(\ref{eq: string action}) and simplifying on an Einstein background. The expression below is used with that qualification. It can also be differentiated as in Eq.~(\ref{eq: dyn metric eom}), since $g^{(0)}+b\,g^{(b)}$ obeys the Einstein equation through first order in $b$. It is given by
\begin{equation}
\begin{aligned}
   T_{ab}^{\rm eff}[g] &:= \frac{1}{2} g_{ab} C^{iejf} C_{i\phantom{g}j}^{\phantom{i}g\phantom{j}h}C^{c\phantom{e}d}_{\phantom{c}e\phantom{d}g} C_{cfdh} - 2 C_{a}^{\phantom{a}ejf} C_{b\phantom{g}j}^{\phantom{b}g\phantom{j}h} C_{cedg}C^{c\phantom{f}d}_{\phantom{c}f\phantom{d}h} - 2 C^{i\phantom{a}jf}_{\phantom{i}a}C_{i\phantom{g}j}^{\phantom{i}g\phantom{j}h} C^{c\phantom{b}d}_{\phantom{c}b\phantom{d}g} C_{cfdh}\\
&\quad- 4 \mathcal P_{ab}^{\phantom{ab}cd}[C_{cedf}C^{eijk} C^{f}_{\phantom{f}ijk}] + 8 \mathcal P_{ab}^{\phantom{ab}cd}[C_{c\phantom{g}e}^{\phantom{c}g\phantom{e}h} C^{iej}_{\phantom{iej}g} C_{idjh}] + C_{aicd} C_{bjef} C_{\phantom{j}g}^{j \phantom{g}cd}C^{igef} \\
&\quad- 2 \nabla_i \nabla_j \left(C^{igjh} C^{c\phantom{a}d}_{\phantom{c}a\phantom{d}g} C_{cbdh} \right) - 2 \nabla_e \nabla_f\left( C_{a\phantom{g}b}^{\phantom{a}g\phantom{b}h} C^{ced}_{\phantom{ced}g} C_{c\phantom{f}dh}^{\phantom{c}f}\right)\\
&\quad - \frac{1}{8} g_{ab} C^{ijef}\, C_{ij}^{\phantom{ij}gh} \, C^{cd}_{\phantom{cd}eg}\, C_{cdfh} + 4 \nabla_{(e}\nabla_{f)}\left(C_a^{\phantom{a}gfh} C^{ced}_{\phantom{ced}g} C_{cbdh} \right) \\
&\quad -2 \nabla_c \nabla_d \left(C_{(a|\phantom{c}ij}^{\phantom{(a|}c} C_{|b)}^{\phantom{|b)}jef} C_{ef}^{\phantom{ef}di}\right) + 2 \mathcal P_{ab}^{\phantom{ab}cd}\left[C_{cgef} C_{d\phantom{g}ij}^{\phantom{d}g}C^{efij} \right]
\end{aligned}
\end{equation}
where
\begin{equation}
\begin{aligned}
    \mathcal R_{ab}[f] &:= \nabla_a \nabla_b f - g_{ab} \Box f + \frac{4}{L^2} g_{ab} f\\
    \mathcal R_{ab}^{\phantom{ab}cd}[f_{cd}] &:= \nabla_c \nabla_{(a} f_{b)}^{\phantom{b)}c} -\frac{1}{2} \Box f_{ab} -\frac{1}{2} g_{ab} \nabla_c \nabla_d f^{cd}\\
    \mathcal P_{ab}^{\phantom{ab}cd}[f_{cd}]&:= \frac{1}{3}\left( \mathcal R_{ab}^{\phantom{ab}cd}[f_{cd}] - \frac{1}{8}\mathcal R_{ab}[f_{cd}g^{cd}]+\frac{5}{2L^2}f_{ab}\right)\, .
\end{aligned}
\end{equation}

\section{Numerical diagnostics and convergence tests}\label{app: conv}

\subsection{Checks of background correction}
\begin{figure}[t]
    \centering
    \includegraphics[width=0.48\textwidth]{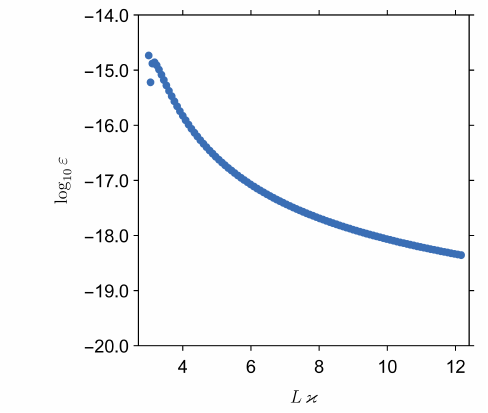}
    \caption{Comparison between the analytic formula Eq.~(\ref{eq: schw area variation}) and the numerical formula Eq.~(\ref{eq: num area variation}) for horizon-area, verifying that the error $\varepsilon =  \left| \frac{\delta A_{{\rm MPAdS}_5,a=0}}{\delta A_{\rm Schwarzschild}}-1\right| $ is at most $10^{-15}$, which is our minimum working precision.}
    \label{fig: area test}
\end{figure}

We provide some basic tests of the accuracy of our determination of the EFT corrections to the MPAdS$_5$ geometry. 
First, we compare with corrections to the Schwarzschild-AdS$_5$ geometry, which can be found analytically to be
\begin{equation}\label{eq: schw metric}
    \dd s^2 = - f(r) \dd t^2 + \frac{\dd r^2}{g(r)}+r^2 \, \dd \Omega_{S^3}^2\, ,
\end{equation}
where
\begin{equation}
\begin{aligned}
    f(r) &= 1- \mu/r^2 + r^2/L^2 + c \, \frac{5\,(L^2+r_H^2)^3}{L^2\, r_H^4\, r^{14}\,(2r_H^2-L^2)}
\Big[ (3L^4 + 5L^2 r_H^2 - 6 r_H^4)\, r^{12}
\\&\qquad+ 24\,(2r_H^2-L^2)\, r_H^{10}\, r^4
+ 16\, L^2 (2r_H^2-L^2)\, r_H^{10}\, r^2
- 9\,(2r_H^2-L^2)(L^2+r_H^2)\, r_H^{12} \Big]\\
    g(r) &= 1 - \mu/r^2 + r^2/L^2 + c \,\frac{5\,(L^2+r_H^2)^3}{L^2\, r_H^4\, r^{14}\,(2r_H^2-L^2)}
\Big[ (3L^4 + 5L^2 r_H^2 - 6 r_H^4)\, r^{12}
\\&\qquad+ 72\,(2r_H^2-L^2)\, r_H^{10}\, r^4
+ 64\, L^2 (2r_H^2-L^2)\, r_H^{10}\, r^2
- 57\,(2r_H^2-L^2)(L^2+r_H^2)\, r_H^{12} \Big]\\
\end{aligned}    
\end{equation}
and $\mu = r_H^2 (L^2 + r_H^2)/L^2$. This correction is in the GCE, so has the same surface gravity $\varkappa = r_H^{-1} + 2 r_H L^{-2}$ as the solution in GR. Note that in this gauge we do not fix the horizon to be at $r=r_H$, so instead the horizon is located at coordinate
\begin{equation}
    r= \tilde r_H = r_H + c\frac{10 (L^2 - 3 r_H^2) (L^2+ r_H^2)^3}{r_H^5 (2 r_H^2 - L^2)}\, ,
\end{equation}
giving the horizon a cross-sectional area of
\begin{equation}\label{eq: schw area variation}
    A_{\rm Schwarzschild} = 2 \pi^2 \tilde r_H^3 = 2 \pi^2 r_H^3 + c\, \frac{60 \pi^2 (L^2 - 3 r_H^2)(L^2 + r_H^2)^3}{r_H^3(2r_H^2 - L^2)}  \,.
\end{equation}
Although the EFT-corrected metric in Eq.~(\ref{eq: schw metric}) can be found in closed form, this does not mean that the $B_i(r)$ which correct the MPAdS$_5$ metric can also be written in closed form when $a=0$, as these metrics are written in different gauges, and the required redefinition of $r$ to convert between these is non-trivial. To verify that these metrics describe the same geometry, we instead find an observable in both which is easy to calculate and compare the numerical value in the two different methods. For this observable, we use the horizon cross-sectional area (note that since we are working in EFT, this is \emph{not} proportional to the entropy). We compare the numerical evaluation of this in the GCE for $a=0$ MPAdS$_5$. In the numerical ensemble, the variation is
\begin{equation}\label{eq: num area variation}
    \delta A_{{\rm MPAdS}_5,a=0}^{\rm num} = 4 L^2 \pi^2 r_H [B_3(0)+3B_4(0)]\, .
\end{equation}
We transform this to the GCE via the standard method as in Section~\ref{sec: num methods}, and then compare with the closed form calculation for Schwarzschild-AdS$_5$ in Figure~\ref{fig: area test}, where we see a clear numerical agreement up to order $10^{-15}$. 

A further test is to check that the residual of all the equations of motion and $\mathcal C$, as defined in Section~\ref{sec: EFT MP Corrections}, are both close to zero. For the numerical run used to generate the plots in Section~\ref{sec: num results}, we find these are both of order $10^{-15}$, providing a strong test of our numerics and of the consistency of our equations.

\subsection{Diagnostic for the coalescing pair}
The structure of the numerical methods of Section~\ref{sec: num methods} already shows that we are going to have arbitrarily large corrections $\delta \omega$ near an EP: when calculating $\delta \omega$, we use the following denominator (which here we have normalised)
\begin{equation}\label{eq: Delta def}
    \Delta := \frac{\mathbf Y_D^T P'(\omega_0) \mathbf Y_0}{\|\mathbf Y_D\| \times \|\mathbf Y_0\|}\, .
\end{equation}
Note that this denominator also appears in numerical expressions we would need to calculate $\partial_{r_H} \omega$, $\partial_{a}\omega$, both of which are expected to have an inverse-square root divergence as we approach the relevant EP. As such, we expect this denominator $\Delta$ to get arbitrarily close to zero. This is a clear diagnostic of the amplified corrections $\delta \omega$ we expect to find near the EP. The quantity $\Delta$ is generally complex and its phase depends on the normalisations of the left and right eigenvectors. We therefore plot $|\Delta|$. Its vanishing is the relevant diagnostic; its magnitude also depends on the scaling of the discretised equations.
\begin{figure}[t]
    \centering
    \includegraphics[width=\textwidth]{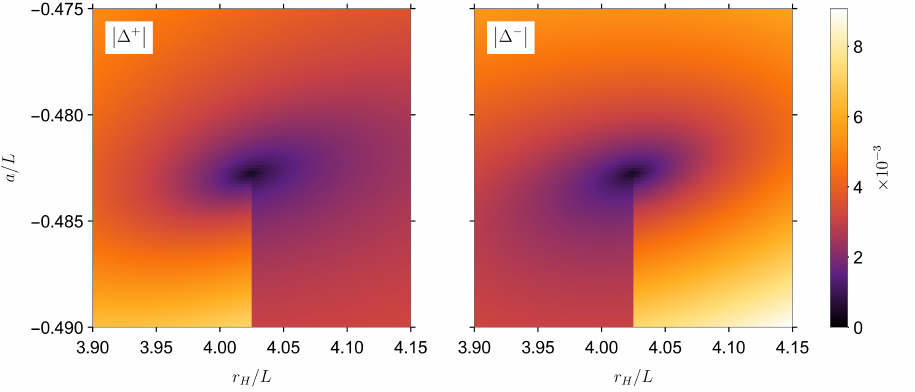}
    \caption{The magnitude $|\Delta|$ of the diagnostic in Eq.~(\ref{eq: Delta def}) for the two $J=0$ branches at $N=25$, near the EP of Figure~\ref{fig: exceptional QNMs}.}
    \label{fig: Delta plot}
\end{figure}

\subsection{Resolution convergence on the sampled grid}
\begin{figure}[t]
    \centering
    \includegraphics[width=\textwidth]{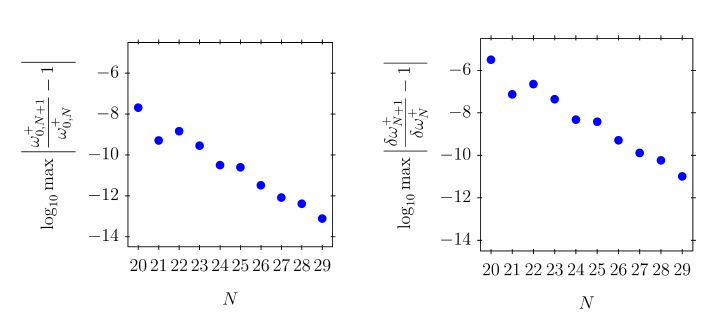}
    \caption{Maximum fractional differences between successive Chebyshev resolutions for $\omega_0^+$ and $\delta\omega^+$ on the $10\times10$ parameter grid specified in the text. These are sampled resolution tests, rather than bounds arbitrarily close to the EP.}
    \label{fig: conv}
\end{figure}

We verify resolution convergence by zooming in close to the EP and testing a sample of points there. We restrict to
\begin{equation}
    4.00\leq\frac{r_H}{L}\leq4.05\,,\qquad
    -0.4832\leq\frac aL\leq-0.4822\, ,
\end{equation}
and calculate $\omega_0^+$ and $\delta\omega^+$ on a $10\times10$ uniformly distributed parameter grid for Chebyshev resolutions $N=20,21,\ldots,30$. We label these as $\omega_{0,N}^+$ and $\delta\omega_N^+$ and plot the logarithm of the maximum fractional change between successive resolutions across the sampled points. The rapid convergence expected of spectral methods is evident in Figure~\ref{fig: conv}. At $N=25$, the displayed successive-resolution differences are below $10^{-8}$ on this sample. Increasing resolution at fixed arithmetic precision does not by itself test roundoff sensitivity or establish a uniform bound as the gap tends to zero.

We can also report that, at $N=25$, the residuals
\begin{equation}
    \frac{\| \mathbf P(\omega)\mathbf Y_0\|}{\|\mathbf P(\omega)\| \, \|\mathbf Y_0\|} \qquad \text{and} \qquad \frac{\| \mathbf P(\omega)^T\mathbf Y_D\|}{\|\mathbf P(\omega)\| \, \|\mathbf Y_D\|}
\end{equation}
in this test are of order $10^{-15}$, verifying that we are indeed solving this matrix problem.

\subsection{Convergence to the EP}
We first verify that our Newton solver to find the EP by varying $(r_H,a)$ converges, and that the corresponding calculation of $\eta_c$ is stable under changing $N$, our minimum precision, and our tolerance for error. 

To more carefully approach the EP, we instead terminate our Newton solver when the step sizes in $r_H /L$ and $a/L$, and the residual $\frac{\| \mathbf P(\omega)\mathbf Y_0\|}{\|\mathbf P(\omega)\| \, \|\mathbf Y_0\|}$ is below a given error tolerance. 

Varying $N = 25, 30, 35$, minimum precision at $15, 20, 25$ digits, and error tolerance of $10^{-20}, 10^{-25}, 10^{-30}$ only produces fractional changes in $\omega_E, (r_H/L)_E, (a/L)_E, L^2 \eta_c, \partial_c(L \varkappa)_E$, and $\partial_c(L \Omega_\psi)_E$ of order $10^{-10}$, verifying that these calculations are numerically stable. \par At the EP, the normalised residual of Eq.~(\ref{eq: Y1 def}), \begin{equation} \frac{\|\mathbf P(\omega_E)\mathbf Y_1+\mathbf P'(\omega_E)\mathbf Y_0\|}{\|\mathbf P(\omega_E)\mathbf Y_1\|+\|\mathbf P'(\omega_E)\mathbf Y_0\|}\,, \end{equation} is below $10^{-21}$ for the numerical settings listed above.

Repeating the approach calculation with the increased resolutions and minimum precisions listed above changes the logarithmic slopes for $\theta = 0, 2\pi$ (which we choose as a representative pair of branches) in Figure~\ref{fig: slopes} by less than $10^{-3}$ and the estimation of $\eta_c$ by a fractional change of less than $10^{-6}$ over $-15<\log \tau<-12$. The limiting slopes remain consistent with $1/2$ and $-1/2$.

\bibliography{references.bib}{}
\bibliographystyle{utphys-modified}

\end{document}